\documentclass[twocolumn,secnumarabic,amssymb, nobibnotes, aps, prb, superscriptaddress]{revtex4-1}
\usepackage[colorlinks=true,linkcolor=blue,citecolor=blue]{hyperref}
\usepackage{graphics}
\usepackage{graphicx}

\usepackage{amsmath}
\usepackage{float}
\usepackage{amsfonts}
\usepackage{mathrsfs}
\usepackage{textcomp}
\usepackage{placeins}

\DeclareGraphicsExtensions{.eps}

\newcommand{\bra}[1]{\langle #1 |}
\newcommand{\ket}[1]{| #1 \rangle}

\begin{document}

\title{Strain engineering of the silicon-vacancy center in diamond}%
\author{Srujan Meesala}
\thanks{These authors contributed equally}
\author{Young-Ik Sohn}%
\thanks{These authors contributed equally}
\affiliation{John A. Paulson School of Engineering and Applied Sciences, Harvard University, 29 Oxford Street, Cambridge, MA 02138, USA}
\author{Benjamin Pingault}
\affiliation{Cavendish Laboratory, University of Cambridge, J. J. Thomson Avenue, Cambridge CB3 0HE, UK}
\author{Linbo Shao}
\affiliation{John A. Paulson School of Engineering and Applied Sciences, Harvard University, 29 Oxford Street, Cambridge, MA 02138, USA}
\author{Haig A. Atikian}
\affiliation{John A. Paulson School of Engineering and Applied Sciences, Harvard University, 29 Oxford Street, Cambridge, MA 02138, USA}
\author{Jeffrey Holzgrafe}
\affiliation{John A. Paulson School of Engineering and Applied Sciences, Harvard University, 29 Oxford Street, Cambridge, MA 02138, USA}
\author{Mustafa G\"undo\u{g}an}
\affiliation{Cavendish Laboratory, University of Cambridge, J. J. Thomson Avenue, Cambridge CB3 0HE, UK}
\author{Camille Stavrakas}
\affiliation{Cavendish Laboratory, University of Cambridge, J. J. Thomson Avenue, Cambridge CB3 0HE, UK}
\author{Alp Sipahigil}
\affiliation{Department of Physics, Harvard University, 17 Oxford Street, Cambridge, MA 02138, USA}
\affiliation{Institute for Quantum Information and Matter and Thomas J. Watson, Sr., Laboratory of Applied Physics, California Institute of Technology, Pasadena, California 91125, USA}
\author{Cleaven Chia}
\affiliation{John A. Paulson School of Engineering and Applied Sciences, Harvard University, 29 Oxford Street, Cambridge, MA 02138, USA}
\author{Michael J. Burek}
\affiliation{John A. Paulson School of Engineering and Applied Sciences, Harvard University, 29 Oxford Street, Cambridge, MA 02138, USA}
\author{Mian Zhang}
\affiliation{John A. Paulson School of Engineering and Applied Sciences, Harvard University, 29 Oxford Street, Cambridge, MA 02138, USA}
\author{Lue Wu}
\affiliation{John A. Paulson School of Engineering and Applied Sciences, Harvard University, 29 Oxford Street, Cambridge, MA 02138, USA}
\author{Jose L. Pacheco}
\affiliation{Sandia National Laboratories, Albuquerque, NM 87185, USA}
\author{John Abraham}
\affiliation{Sandia National Laboratories, Albuquerque, NM 87185, USA}
\author{Edward Bielejec}
\affiliation{Sandia National Laboratories, Albuquerque, NM 87185, USA}
\author{Mikhail D. Lukin}
\affiliation{Department of Physics, Harvard University, 17 Oxford Street, Cambridge, MA 02138, USA}
\author{Mete Atat\"ure}
\affiliation{Cavendish Laboratory, University of Cambridge, J. J. Thomson Avenue, Cambridge CB3 0HE, UK}
\author{Marko Lon\v{c}ar}
\affiliation{John A. Paulson School of Engineering and Applied Sciences, Harvard University, 29 Oxford Street, Cambridge, MA 02138, USA}

\begin{abstract}
We control the electronic structure of the silicon-vacancy (SiV) color-center in diamond by changing its static strain environment with a nano-electro-mechanical system. This allows deterministic and local tuning of SiV optical and spin transition frequencies over a wide range, an essential step towards multi-qubit networks. In the process, we infer the strain Hamiltonian of the SiV revealing large strain susceptibilities of order 1 PHz/strain for the electronic orbital states. We identify regimes where the spin-orbit interaction results in a large strain suseptibility of order 100 THz/strain for spin transitions, and propose an experiment where the SiV spin is strongly coupled to a nanomechanical resonator. 
\end{abstract}
\maketitle

\section{Introduction}
Solid state emitters such as color-centers and epitaxially grown quantum dots provide both electronic spin qubits and coherent optical transitions, and are optically accessible quantum memories. They can therefore serve as building blocks of a quantum network composed of nodes in which information is stored in spin qubits and interactions between nodes are mediated by photons\cite{kimble_quantum_2008, bernien_heralded_2013, delteil2016generation, stockill_phase_2017}. However, due to the effects of their complex solid state environment, most quantum emitters do not simultaneously provide long coherence time for the memory, and favorable optical properties such as bright, spectrally stable emission. The negatively charged silicon vacancy center in diamond (SiV$^-$, hereafter simply referred to as SiV) has been recently identified as a system that can overcome these limitations, since it provides excellent optical and spin properties simultaneously. Its dominant zero-phonon-line (ZPL) emission and stable optical transition frequencies resulting from its inversion symmetry\cite{gali2013ab, muller_optical_2014, sipahigil_indistinguishable_2014} have recently been used to realize single-photon switching\cite{sipahigil_integrated_2016} and a fibre-coupled coherent single-photon source\cite{burek_fiber-coupled_2017} in a nanophotonic platform. Further, recent demonstrations of microwave\cite{pingault_coherent_2017} and all-optical\cite{becker2017all} control of its electronic spin, as well as long ($\sim$10 ms) spin coherence times at mK temperatures\cite{sukachev2017silicon}, when electron-phonon processes in the center are suppressed,\cite{jahnke_electronphonon_2015, pingault_coherent_2017} make the SiV a good spin qubit.

Scaling up these demonstrations to multi-qubit networks requires local tunability of individual emitters, as well as the realization of strong interactions between them. In this work, we control local strain in the SiV environment using a nano-electro-mechanical system (NEMS), and show wide tunability for both optical and spin transition frequencies. In particular, we demonstrate hundreds of GHz of optical tuning, sufficient to achieve spectrally identical emitters for photon-mediated entanglement\cite{kimble_quantum_2008, bernien_heralded_2013}. Further, we characterize the strain Hamiltonian of the SiV and measure high strain susceptibilities for both the electronic and spin levels. Building on this strain response, we discuss a scheme to realize strong coupling of the SiV spin to coherent phonons in GHz frequency nanomechanical resonators. While phonons have been proposed as quantum transducers for qubits,\cite{wallquist_hybrid_2009,rabl_quantum_2010} experiments with solid-state spins have been limited to the classical regime of large displacement amplitudes driving their internal levels\cite{arcizet_single_2011, kolkowitz_coherent_2012, ovartchaiyapong_dynamic_2014, teissier_2014, barfuss_strong_2015, macquarrie_coherent_2015, golter_optomechanical_2016, golter_saw_2016, meesala_enhanced_2016}. The high strain susceptibility of the SiV ground states can enable MHz spin-phonon coupling rates in existing nanomechanical resonators. Such a spin-phonon interface can enable quantum gates between spins akin to those in ion traps\cite{cirac-zoller_1995, molmer-sorensen_1999, leibfried_experimental_2003}, and interfaces with disparate qubits\cite{schuetz_universal_2015,aref_2015}.

\section{Strain tuning of optical transitions}

\begin{figure}[ht!]
\includegraphics[width=\columnwidth]{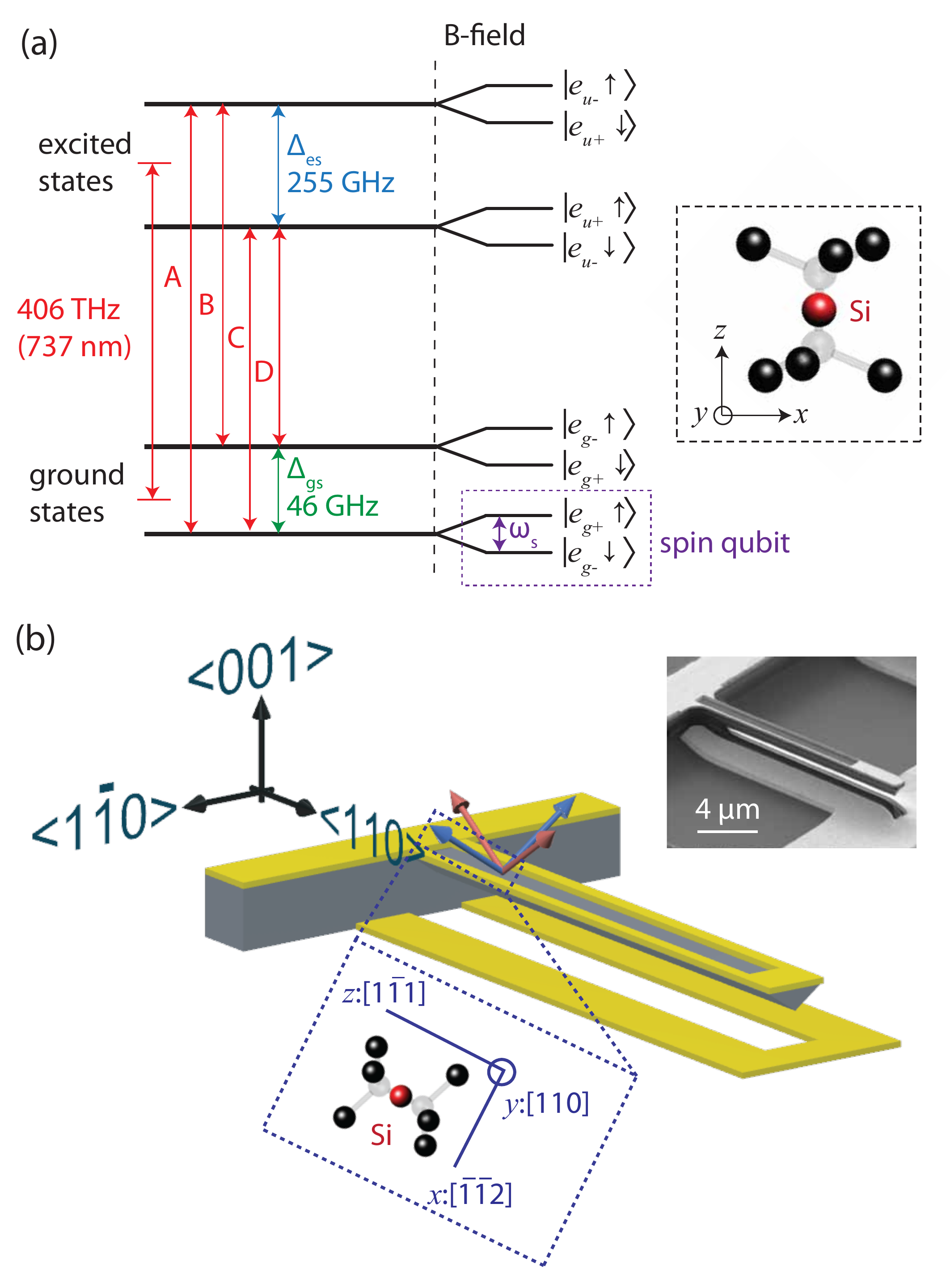}
\caption{(a) Electronic level structure of the SiV center (molecular structure shown in inset) at zero strain showing ground and excited manifolds with spin-orbit eigenstates. The four optical transitions A, B, C, and D at zero magnetic field, and splittings between orbital branches in the ground state (GS) and excited state (ES), $\Delta_{\mathrm{gs}}$ and $\Delta_{\mathrm{es}}$ respectively are indicated. In the presence of a magnetic field, each orbital branch splits into two Zeeman sublevels. A spin-qubit can be defined in the sublevels of the lower orbital branch in the GS. (b) Schematic of the diamond cantilever device and surrounding electrodes with a corresponding scanning electron microscope (SEM) image in the inset. Diamond crystal axes relative to the cantilever orientation are shown. Four possible orientations of the highest symmetry axis of an SiV are indicated by the four arrows above the cantilever. Under application of strain, these can be grouped into axial (red) and transverse (blue) orientations. Molecular structure of a transverse-orientation SiV as viewed in the plane normal to the cantilever axis is shown below, and crystal axes that define the internal co-ordinate frame of the color center are indicated. The $z$-axis is the highest symmetry axis, which defines the orientation of the SiV. }
\label{fig1}
\end{figure}

The SiV center is an interstitial point defect in which a silicon atom is positioned midway between two adjacent missing carbon atoms in the diamond lattice as depicted in the inset of Fig. \ref{fig1}(a). Its electronic level structure at zero strain is shown in Fig. \ref{fig1}(a). The optical ground state (GS) and excited state (ES) each contain two distinct electronic configurations shown by the bold horizontal lines. Physically, each of the two branches in the GS and ES corresponds to the occupation of a specific $E$-symmetry orbital by an unpaired hole.\cite{HeppThesis} At zero magnetic field, the degeneracy of these orbitals is broken by spin-orbit (SO) coupling leading to frequency splittings $\Delta_{\mathrm{gs}}$ = 46 GHz, and $\Delta_{\mathrm{es}}$ = 255 GHz respectively. Due to inversion symmetry of the defect about the Si atom, the wavefunctions of these orbitals can be classified according to their parity with respect to this inversion center.\cite{gali2013ab, HeppThesis} Thus, the GS configurations correspond to the presence of the unpaired hole in one of the even-parity orbitals $e_{g+}, e_{g-}$, while the ES configurations have this hole in one of the odd-parity orbitals $e_{u+}, e_{u-}$. Here the subscripts $g$, $u$ refer to even ($gerade$) and odd ($ungerade$) parity respectively, and $+$, $-$ refer to the orbital angular momentum projecton $l_Z$. This specific level structure gives rise to four distinct optical transitions in the ZPL indicated by A, B, C, D in Fig. \ref{fig1}(a). Upon application of a magnetic field, degeneracy between the SO eigenstates is further broken to reveal two sub-levels within each orbital branch corresponding to different spin states of the unpaired hole ($S= 1/2$). In this manner, a spin-qubit can be defined on the two sublevels of the lowest orbital branch in the ground state.

To control local strain in the environment of the SiV center, we use a diamond cantilever, schematically shown in Fig. \ref{fig1}(b). Electrodes are fabricated, one on top of the cantilever, and another on the substrate below the cantilever to form a capacitive actuator. By applying a specific DC voltage to these electrodes, we can deflect the cantilever to achieve a desired amount of static strain at the SiV site. The fabrication procedure based on angled etching of diamond \cite{atikian2017freestanding,burek_free-standing_2012} and device design are discussed in detail elsewhere \cite{sohn2017controlling}. The diamond sample with cantilever NEMS is maintained at 4 K in a Janis, ST-500 continuous-flow liquid helium cryostat. We perform optical spectroscopy on SiVs inside the cantilever via resonant laser excitation of the transitions shown in Fig. \ref{fig1}(a). Mapping the response of these transitions as a function of voltage applied to the device allows us to study the strain response of the SiV electronic structure.

The diamond samples used in our study have a $[001]$-oriented top surface, and the long axis of the cantilever is oriented along the $[110]$ direction. There are four possible equivalent orientations of SiVs - $[111]$, $[\bar{1}\bar{1}1]$, $[1\bar{1}1]$, $[\bar{1}11]$ - in a diamond crystal, indicated by the four arrows above the cantilever in Fig. \ref{fig1}(b). Since the cantilever primarily achieves uniaxial strain directed along $[110]$, this breaks the equivalence of the four orientations, and leads to two classes indicated by the blue and red colored arrows in Fig. \ref{fig1}(b). The blue SiVs, oriented perpendicular to the cantilever long-axis, predominantly experience uniaxial strain along their internal $y$-axis (see inset of Fig. \ref{fig1}(b)). On the other hand, the red SiVs are not orthogonal to the cantilever long-axis, and experience a non-trivial strain tensor, which includes significant strain along their internal $z$-axis. For simplicity, we refer to blue SiVs as `transverse-orientation' SiVs, and red SiVs as `axial-orientation' SiVs. This nomenclature is used with the understanding that it is specific to the situation of predominantly $[110]$ uniaxial strain applied with our cantilevers.

\begin{figure}[ht!]
\includegraphics[width=\columnwidth]{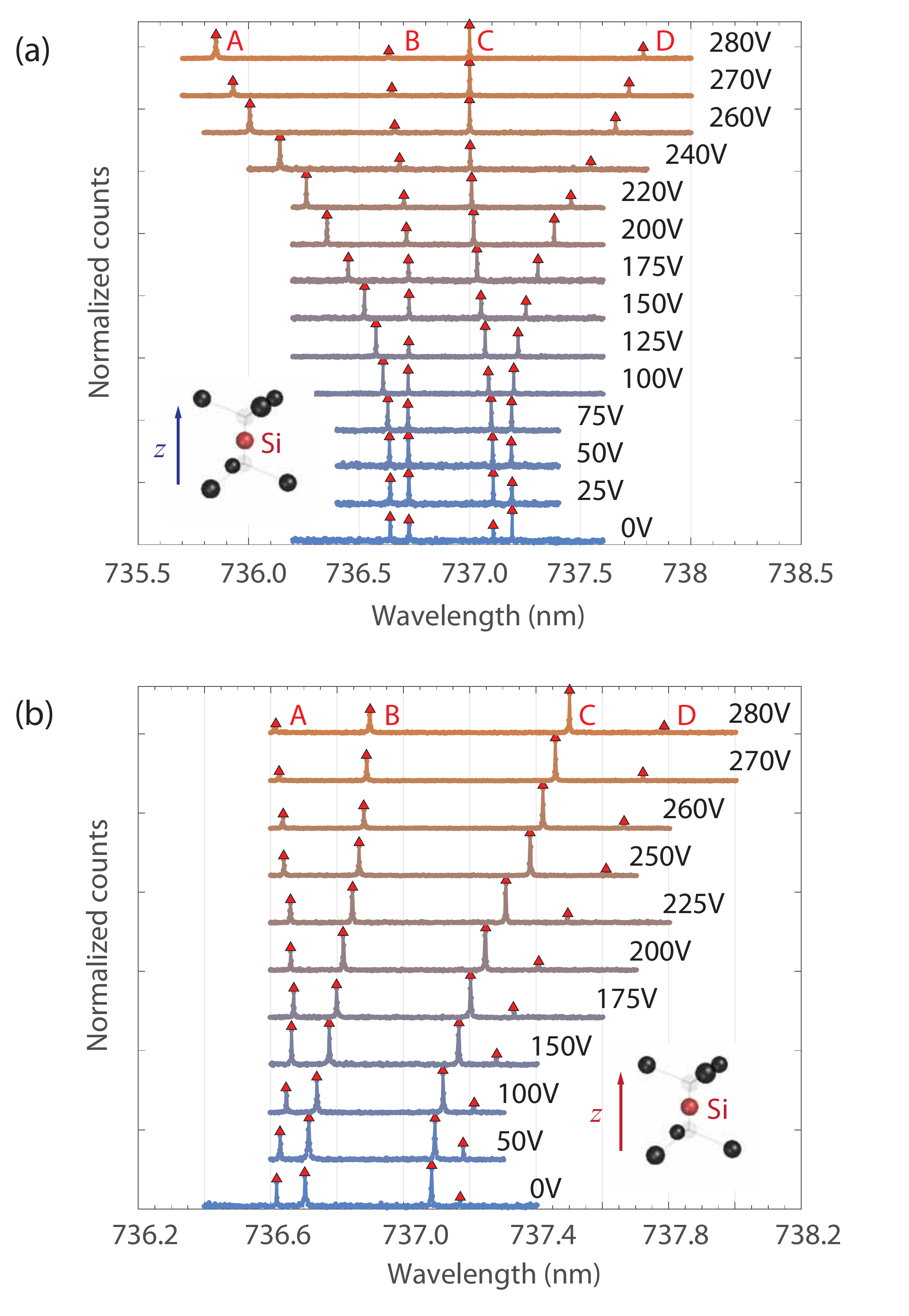}
\caption{Tuning of optical transitions of (a) transverse-orientation SiV (red in Fig. 1(b)), and (b) axial orientation SiV (blue in Fig. 1(b)). Voltage applied to the device is indicated next to each spectrum.}
\label{fig2}
\end{figure}

Two distinct strain-tuning behaviors correlated with SiV orientation are observed as shown in Fig. \ref{fig2}. Orientation of SiVs in the cantilever is inferred from polarization-dependence of their optical transitions at zero strain.\cite{HeppThesis} With gradually increasing strain, transverse-orientation SiVs show an increasing separation between the A and D transitions with relatively small shifts in the B and C transitions as seen in Fig. \ref{fig2}(a). This behavior has been observed on a previous experiment with an ensemble of SiVs.\cite{sternschulte_1.681-ev_1994} On the other hand, axial-orientation SiVs show a more complex tuning behavior in which all transitions shift as seen in Fig. \ref{fig2}(b).

In the context of photon-mediated entanglement of emitters, typically, photons emitted in the C line, the brightest and narrowest linewidth transition are of interest\cite{sipahigil_integrated_2016}. Upon comparing Figs. \ref{fig2}(a) and (b), we note that this transition is significantly more responsive for axial-orientation SiVs. Particularly in Fig. \ref{fig2}(b), we achieve tuning of the C transition wavelength by 0.3 nm (150 GHz), approximately 10 times the typical inhomogeneity in optical transition frequencies of SiV centers.\cite{muller_optical_2014,evans_narrow-linewidth_2016} Thus, NEMS-based strain control can be used to deterministically tune multiple on-chip or distant emitters to a set optical wavelength. In particular, integration of this NEMS-based strain-tuning with existing diamond nanophotonic devices\cite{sipahigil_integrated_2016, burek_fiber-coupled_2017, bhaskar_gev_2017, zhang_2017, Burek:2014bj} can enable scalable on-chip entanglement and widely tunable single photon sources. Besides static tuning of emitters, dynamic control of the voltage applied to the NEMS can be used to counteract slow spectral diffusion, and stabilize optical transition frequencies\cite{acosta_dynamic_2012}.

\section{Effect of strain on electronic structure}
\begin{figure}[ht!]
\includegraphics[width=\columnwidth]{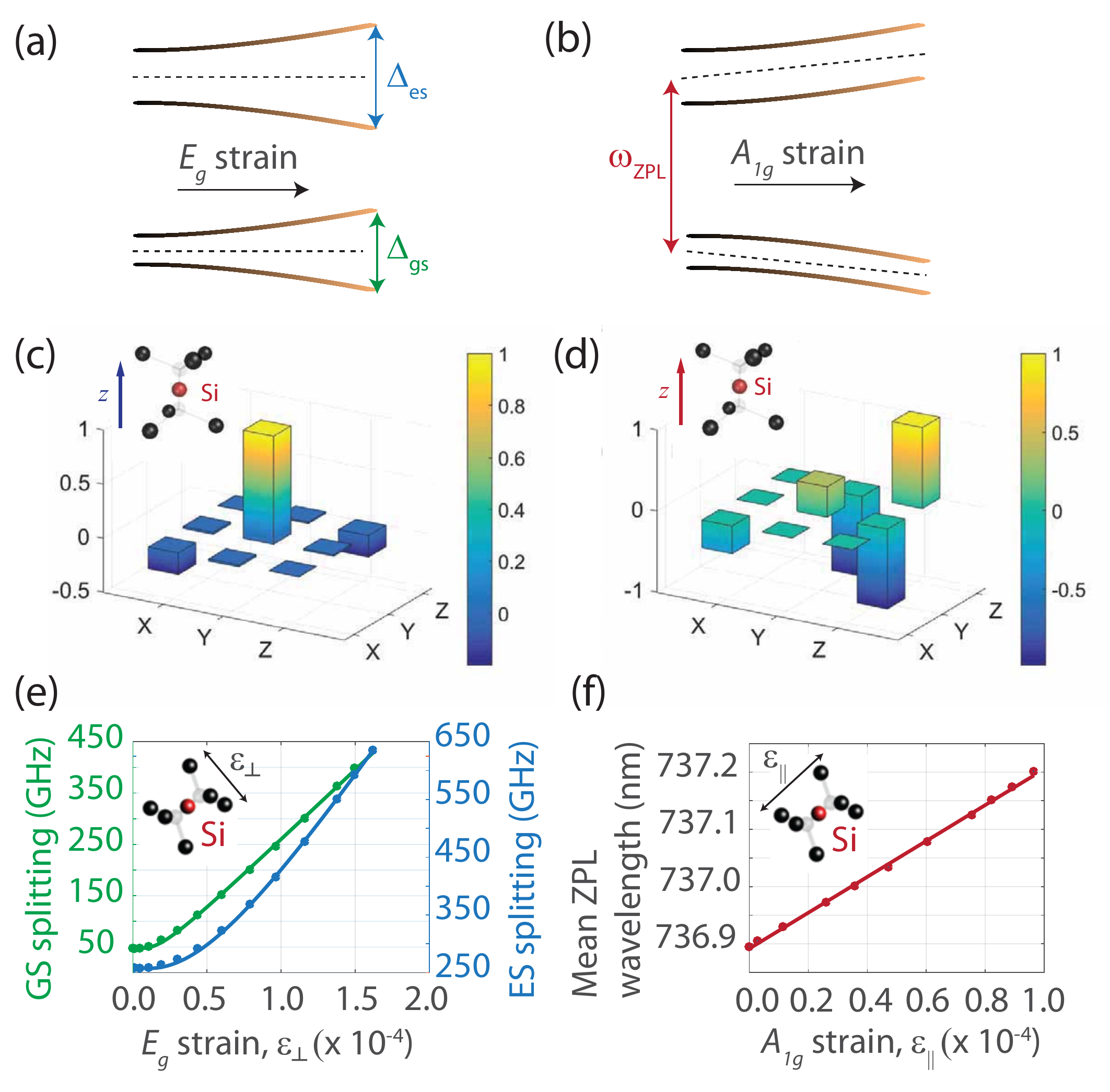}
\caption{(a) Dominant effect of $E_g$-strain on the electronic levels of the SiV. (b) Dominant effect of $A_{1g}$-strain on the electronic levels of the SiV. (c) Normalized strain-tensor components experienced by transverse-orientation SiV (red in Fig. 1(b)), and (d) axial orientation SiV (blue in Fig. 1(b)) in the SiV co-ordinate frame upon deflection of the cantilever. (e) Variation in orbital splittings within GS (solid green squares) and ES (open blue circles) upon application of $E_g$-strain. Data points are extracted from the optical spectra in Fig. 2(a). Solid curves are fits to theory in text. (f) Tuning of mean optical wavelength with $A_{1g}$ strain. Data points are extracted from the optical spectra in Fig. 2(b). Solid line is a linear fit as predicted by theory in text.}
\label{fig3}
\end{figure}
Following previous work on point defects,\cite{hughes_uniaxial_1967, maze_properties_2011, HeppThesis} we employ group theory to explain the effect of strain on the SiV electronic levels, and extract the susceptibilities for various strain components.

\subsection{Strain Hamiltonian}\label{strain.theory}
In this section, we describe the strain Hamiltonian of the SiV center, and summarize the physical effects of various modes of deformation on the orbital wavefunctions. A more detailed group-theoretic discussion of the results in this section is provided in Appendix \ref{group.theory} and in Ref. \cite{HeppThesis}. Based on the symmetries of the orbital wavefunctions, it can be shown that the effects of strain on the GS ($e_g$) and ES ($e_u$) manifolds are independent and identical in form. For either manifold, the strain Hamiltonian in the basis of $\{\ket{e_x\downarrow}, \ket{e_x\uparrow}, \ket{e_y\downarrow}, \ket{e_y\uparrow}\}$ states (pure orbitals unmixed by SO coupling as defined in \cite{HeppThesis}) is given by

\begin{equation}
\mathbb{H}^{\text{strain}} = \begin{bmatrix}
\epsilon_{A_{1g}}-\epsilon_{E_{gx}} & \epsilon_{E_{gy}}  \\
\epsilon_{E_{gy}} & \epsilon_{A_{1g}}+\epsilon_{E_{gx}}  \\
\end{bmatrix}\otimes \mathbb{I}_{2}
\label{eq:Hstr.spin}
\end{equation}

The spin part of the wavefunction is associated with an identity matrix in Eq. (\ref{eq:Hstr.spin}) because lattice deformation predominantly perturbs the Coulomb energy of the orbitals, which is independent of the spin character. Each $\epsilon_r$ is a linear combination of strain components $\epsilon_{ij}$, and corresponds to specific symmetries indicated by the subscript $r$.

\begin{eqnarray}
\epsilon_{A_{1g}} & = & t_\perp(\epsilon_{xx}+\epsilon_{yy}) + t_\parallel\epsilon_{zz} \nonumber\\
\epsilon_{E_{gx}} & = & d(\epsilon_{xx}-\epsilon_{yy}) + f\epsilon_{zx} \label{eq:strain.siv.basis}\\
\epsilon_{E_{gy}} & = & -2d\epsilon_{xy} + f\epsilon_{yz} \nonumber
\end{eqnarray}

Here $t_\perp, t_\parallel, d, f$ are the four strain-susceptibility parameters that completely describe the strain-response of the $\{\ket{e_x},\ket{e_y}\}$ states. These parameters have different numerical values in the GS and ES manifolds. From the Hamiltonian \ref{eq:Hstr.spin}, we see that $E_{gx}$ and $E_{gy}$ strain cause mixing and relative shifts between orbitals, and modify the orbital splittings within the GS and ES manifolds as depicted in Fig. \ref{fig3}(a). On the other hand, $A_{1g}$ strain leads to a uniform or common-mode shift of the GS and ES manifolds, and only shifts the mean ZPL frequency as depicted in Fig. \ref{fig3}(b). 


By decomposing the strain applied in our experiment into $A_{1g}$ and $E_g$ components, we can confirm the observations on tuning of transverse- and axial-orientation SiVs in Fig. \ref{fig2}. Strain tensors for transverse- and axial-orientations of emitters obtained from finite element method (FEM) simulations are plotted in Figs. \ref{fig3}(c), (d) respectively. As expected from the cantilever geometry in Fig. \ref{fig1}(a), transverse-orientation SiVs predominantly experience $\epsilon_{yy}$ and hence an $E_g$ deformation. The $E_g$-strain response predicted in Fig. \ref{fig3}(a) leads to the strain-tuning of mainly A and D transitions seen in Fig. \ref{fig2}(a). On the other hand, axial-orientation SiVs experience both $\epsilon_{zz}$ and $\epsilon_{yz}$ as shown in Fig. \ref{fig3}(d), which leads to simultaneous $E_g$ and $A_{1g}$ deformations. Indeed, a combination of the strain responses in Figs. \ref{fig3}(a), (b) qualitatively explains the strain-tuning behavior of the transitions in Fig. \ref{fig2}(b).

\subsection{Estimation of strain-susceptibilities}\label{strain.fitting}
We now quantitatively fit the results in Fig. \ref{fig2} with the above strain response model. Adding SO coupling ($\mathbb{H}^{\text{SO}} = -{\lambda_{SO}}L_zS_z$) to the strain Hamiltonian in Eq. \ref{eq:Hstr.spin}, we get the following total Hamiltonian in the $\{\ket{e_x}, \ket{e_y}\} \otimes \{\ket{\uparrow}, \ket{\downarrow}\}$ basis.\cite{HeppThesis}

\begin{widetext}
\begin{equation}
\mathbb{H}^{\text{total}} = \left[ {\begin{array}{cccc}  \epsilon_{A_{1g}}-\epsilon_{E_{gx}} & 0 & \epsilon_{E_{gy}}-i\lambda_{SO}/2 & 0 \\  0 & \epsilon_{A_{1g}}-\epsilon_{E_{gx}} & 0 & \epsilon_{E_{gy}}+i\lambda_{SO}/2 \\ \epsilon_{E_{gy}}+i\lambda_{SO}/2 & 0 & \epsilon_{A_{1g}}+\epsilon_{E_{gx}} & 0 \\ 0 & \epsilon_{E_{gy}}-i\lambda_{SO}/2 & 0 & \epsilon_{A_{1g}}+\epsilon_{E_{gx}}\\ \end{array} } \right]
\label{eq:H.SO.strain}
\end{equation}
\end{widetext}

Here, $\lambda_{SO}$ is the SO coupling strength within each manifold - 46 GHz for the GS, and 255 GHz for the ES. Diagonalization of this Hamiltonian gives two distinct eigenvalues

\begin{eqnarray}
E_1 & = & \alpha-\frac{1}{2}\sqrt{\lambda_{SO}^2+4(\epsilon_{E_{gx}}^2+\epsilon_{E_{gy}}^2)} \nonumber \\
E_2 & = & \alpha+\frac{1}{2}\sqrt{\lambda_{SO}^2+4(\epsilon_{E_{gx}}^2+\epsilon_{E_{gy}}^2)}
\label{eq:eigenenergy}
\end{eqnarray}

Each of these corresponds to doubly spin-degenerate eigenstates in the absence of an external magnetic field. Noting that Eqs. (\ref{eq:eigenenergy}) are valid within both GS and ES manifolds, but with different strain susceptibilities, we obtain the following quantities that can be directly extracted from the optical spectra in Fig. \ref{fig2}.

\begin{widetext}
\begin{eqnarray}
\Delta_{\text{ZPL}} &=& \Delta_{\text{ZPL},0}+\left(t_{\parallel,u}-t_{\parallel,g}\right)\epsilon_{zz} +\left(t_{\perp,u}-t_{\perp,g}\right)(\epsilon_{xx}+\epsilon_{yy}) \label{eq:fitmodel1}\\
\Delta_{\mathrm{gs}} &=& \sqrt{\lambda_{SO,g}^2 + 4\left[d_g(\epsilon_{xx}-\epsilon_{yy})+f_g\epsilon_{yz}\right]^2+4\left[-2d_g\epsilon_{xy}+f_g\epsilon_{zx}\right]^2} \label{eq:fitmodel2}\\
\Delta_{\mathrm{es}} &=& \sqrt{\lambda_{SO,u}^2 + 4\left[d_u(\epsilon_{xx}-\epsilon_{yy})+f_u\epsilon_{yz}\right]^2+4\left[-2d_u\epsilon_{xy}+f_u\epsilon_{zx}\right]^2} \label{eq:fitmodel3}
\end{eqnarray}
\end{widetext}

Here, the subscript $g(u)$ refers to the GS (ES) manifold. $\Delta_{\text{ZPL}}$  is the mean ZPL frequency, and $\Delta_{\mathrm{gs}}$, $\Delta_{\mathrm{es}}$ are the GS and ES orbital splittings respectively. $\Delta_{\text{ZPL},0}$ is the mean ZPL frequency at zero strain. Extracting all three frequencies in Eqs. (\ref{eq:fitmodel1}-\ref{eq:fitmodel3}) as a function of strain from the optical spectra measured in Fig. \ref{fig2}, we fit them to the above model in Figs. \ref{fig3}(c), (d), and estimate the strain-susceptibilities. The fitting procedure described in detail in Appendix \ref{strain.susc.fit} gives us

\begin{eqnarray}
\left(t_{\parallel,u}-t_{\parallel,g}\right) & = & -1.7\,\text{PHz/strain} \nonumber\\
\left(t_{\perp,u}-t_{\perp,g}\right) & = & 0.078\,\text{PHz/strain} \nonumber\\
d_g & = & 1.3\,\text{PHz/strain} \nonumber\\
d_u & = & 1.8\,\text{PHz/strain} \nonumber\\
f_g & = &-0.25\,\text{THz/strain} \nonumber\\
f_u & = &-0.72\,\text{THz/strain} \label{eq:str.susc.vals}
\end{eqnarray}

We note that these values are subject to errors arising from (i) imprecision in SiV depth from the diamond surface (10\% from SRIM calculations, and in practice, higher due to ion-channeling effects), and (ii) due to the fact that the device geometry cannot be replicated exactly in FEM simulations for strain estimation. In particular, the values $f$ and $t_{\perp}$ are subject to higher error, since the $E_g$ and $A_{1g}$ responses are mostly dominated by the numerically larger susceptibilities $d$ and $t_{\parallel}$ respectively.

\section{Controlling electron-phonon processes}
\begin{figure}[b!]
\includegraphics[width=\columnwidth]{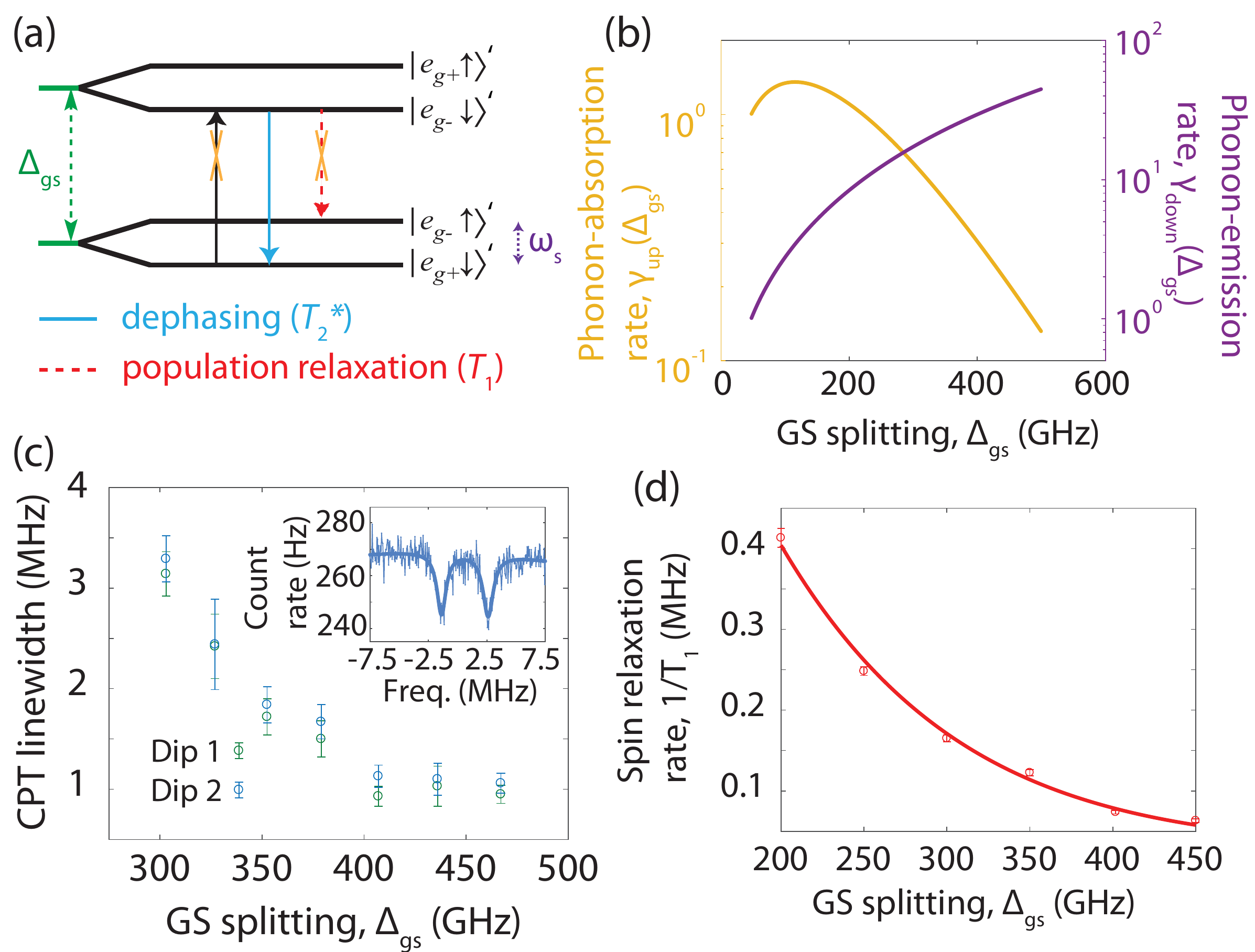}
\caption{(a) Illustration of dephasing and population decay processes for spin qubit. Blue arrow shows a spin-conserving transition responsible for dephasing. Red arrow shows a spin-flipping transition driving decay from $\ket{e_{g+}\downarrow}^\prime$ to $\ket{e_{g-}\uparrow}^\prime$. Processes suppressed at high strain are crossed out. (b) Calculated rates for spin-conserving upward and downward phonon processes. Both rates are normalized to their values at zero strain. (c) Reduction in CPT linewidth with increasing GS splitting $\Delta_{\mathrm{gs}}$. Inset shows an example of a CPT spectrum taken at $\Delta_{\mathrm{gs}}$ = 460 GHz. The two resonances in the spectrum are due to the presence of a neighboring nuclear spin \cite{sohn2017controlling}. Linewidths of both are plotted and indicated as Dip 1 and Dip 2 in the main plot. (d) Reduction in spin relaxation rate ($1/T_1$) with increasing GS splitting $\Delta_{\mathrm{gs}}$ as extracted from pump-probe measurements. Solid line is a fit to theory in Appendix \ref{spin.T1}.}
\label{fig5}
\end{figure}

At 4 K, dephasing and population relaxation of the SiV spin qubit defined with the $\ket{e_{g+}\downarrow}^\prime$, $\ket{e_{g-}\uparrow}^\prime$ states ($^\prime$ denoting modified SO eigenstates due to strain) is known to be dominated by electron-phonon processes shown in Fig. \ref{fig5}(a)\cite{jahnke_electronphonon_2015, pingault_coherent_2017}. In accordance with our observations on response to static $E_g$-strain in the previous section, we expect that AC strain generated by thermal $E_g$-phonons at frequency $\Delta_\mathrm{gs} < k_B T/h$ is capable of driving the GS orbital transitions. Since we can tune the splitting $\Delta_\mathrm{gs}$ by applying static $E_g$-strain with our device, we have control over these electron-phonon processes, and can engineer the relaxation rates of spin qubit. In particular, by making $\Delta_\mathrm{gs} \gg  k_B T/h$, we have shown that spin coherence can be improved significantly.\cite{sohn2017controlling} Here, we elucidate the physical mechanisms behind such improvement in spin properties with strain control.

When a thermal phonon randomly excites the SiV center from the spin qubit manifold to the upper orbital branch, say from $\ket{e_{g+}\downarrow}^\prime$ to $\ket{e_{g-}\downarrow}^\prime$ as shown by the blue upward arrow in Fig. \ref{fig5}(a), the energy of the $\downarrow$ projection of the spin qubit suddenly changes by an amount $h\Delta_{\mathrm{gs}}$. After some time in the upper branch, the system randomly relaxes back to the lower manifold through spontaneous emission of a phonon as shown by the blue downward arrow in Fig. \ref{fig5}(a). In this process, the spin projection is conserved, since phonons predominantly flip only the orbital character. However, a random phase is acquired between the $\downarrow$ and $\uparrow$ projections of the spin qubit due to phonon absorption and emission, as well as faster precession in the upper manifold. The dephasing rate is determined by the upward phonon transition rate $\gamma_{\mathrm{up}}(\Delta_{\mathrm{gs}})$. Both this rate and the downward transition rate $\gamma_{\mathrm{down}}(\Delta_{\mathrm{gs}})$ can be calculated from Fermi's golden rule and are given by -

\begin{align}
\gamma_{\mathrm{up}}(\Delta_{\mathrm{gs}}) &= 2\pi\chi\rho\Delta_{\mathrm{gs}}^3 n_{th}(\Delta_{\mathrm{gs}})
\label{eq:gammaup.final}\\
\gamma_{\mathrm{down}}(\Delta_{\mathrm{gs}}) &= 2\pi\chi\rho\Delta_{\mathrm{gs}}^3 (n_{th}(\Delta_{\mathrm{gs}})+1)
\label{eq:gammadown.final}
\end{align}
where $\chi$ is a constant that encapsulates averaged interaction over all phonon modes and polarizations and $n_{th}(\nu)$ is the Bose-Einstein distribution. It is instructive to view these rates as a product of the phonon density of states (DOS) and the occupation of phonon modes. In the above expressions, the first part $2\pi\chi\rho\Delta_{\mathrm{gs}}^3$ contains the bulk DOS of phonons, which scales as $\sim\Delta_\mathrm{gs}^2$. On the other hand, $n_{th}(\nu)$ is the number of thermal phonons in each mode. Note that the +1 term in the downward rate in Eq. (\ref{eq:gammadown.final}) corresponds to spontaneous emission of a phonon, a process that is independent of temperature. 

Fig. \ref{fig5}(b) shows the theoretically predicted behavior of upward and downward rates as a function of $\Delta_{\mathrm{gs}}$ at temperature $T=4$\,K. Here, we calculate both transition rates with corrected exponent in Eqs. (\ref{eq:gammaup.final}) and (\ref{eq:gammadown.final}), approximately 1.9 rather than 3, to take into account the geometric factor associated with the cantilever\cite{sohn2017controlling}. We observe that the upward rate shows a non-monotonic behavior, approaching its maximum value around $h\Delta_{\mathrm{gs}} \sim k_BT$. In the $h\Delta_{\mathrm{gs}} < k_BT$ regime, the increasing DOS term dominates, and causes $\gamma_{\mathrm{up}}$ to increase. However, when $h\Delta_{\mathrm{gs}} \gg k_BT$, thermal occupation of the modes is approximated by Boltzmann distribution $n_{th}(\Delta_{\mathrm{gs}}) = \mathrm{exp}\left(-\frac{h\Delta_{\mathrm{gs}}}{k_BT}\right)$, and this exponential roll-off dominates the polynomially increasing DOS. Therefore, $\gamma_{\mathrm{up}}$ decreases exponentially, when sufficiently high strain is applied. In contrast, the downward rate monotonically increases with the GS-splitting, because it is dominated by the spontaneous emission rate, which simply increases polynomially with the DOS. Fig. \ref{fig5}(c) shows experimentally measured improvement of spin coherence using coherent population trapping (CPT) in this high strain regime\cite{sohn2017controlling}. Above $\Delta_\mathrm{gs}$ of 400 GHz, the dephasing rate saturates, indicating a secondary dephasing mechanism such as the $^{13}$C nuclear spin bath in diamond. Our data is supported by similar $1/T_2^*$ measured at 100 mK where the thermal occupation of relevant phonon modes is negligible\cite{sukachev2017silicon}. 


Population decay or longitudinal relaxation of the spin qubit shown by the red arrows in Fig. \ref{fig5}(a) is driven by spin-flipping phonon transitions, which occur with a small probability due to perturbative mixing of spin projections. A detailed analysis of various decay channels is presented in Appendix \ref{spin.T1}. At high strain, it can be shown that the decay rate is approximately $4\left(d_{g, \mathrm{flip}}/d_g\right)^2\gamma_\mathrm{up}$, where $d_{g, \mathrm{flip}}$ is the strain susceptibility for a spin-flipping transition such as $\ket{e_{g+}\downarrow}^\prime \rightarrow \ket{e_{g+}\uparrow}^\prime$. Thus it is a fraction of the spin-conserving transition rate $\gamma_\mathrm{up}$ shown in Eq. \ref{eq:gammaup.final}. The factor $d_{g, \mathrm{flip}}/d_g$ scales as $\sim 1/\Delta_\mathrm{gs}$ according to first order perturbation theory. As a result, we expect exponential decrease in the population decay rate with a different polynomial pre-factor compared to the spin decoherence rate. Fig. \ref{fig5}(d) shows this decreasing trend with increasing $\Delta_\mathrm{gs}$ fit to this two-phonon relaxation model.

\section{Strain response of spin transition}\label{ss.susceptibility}

\begin{figure}[ht!]
\includegraphics[width=\columnwidth]{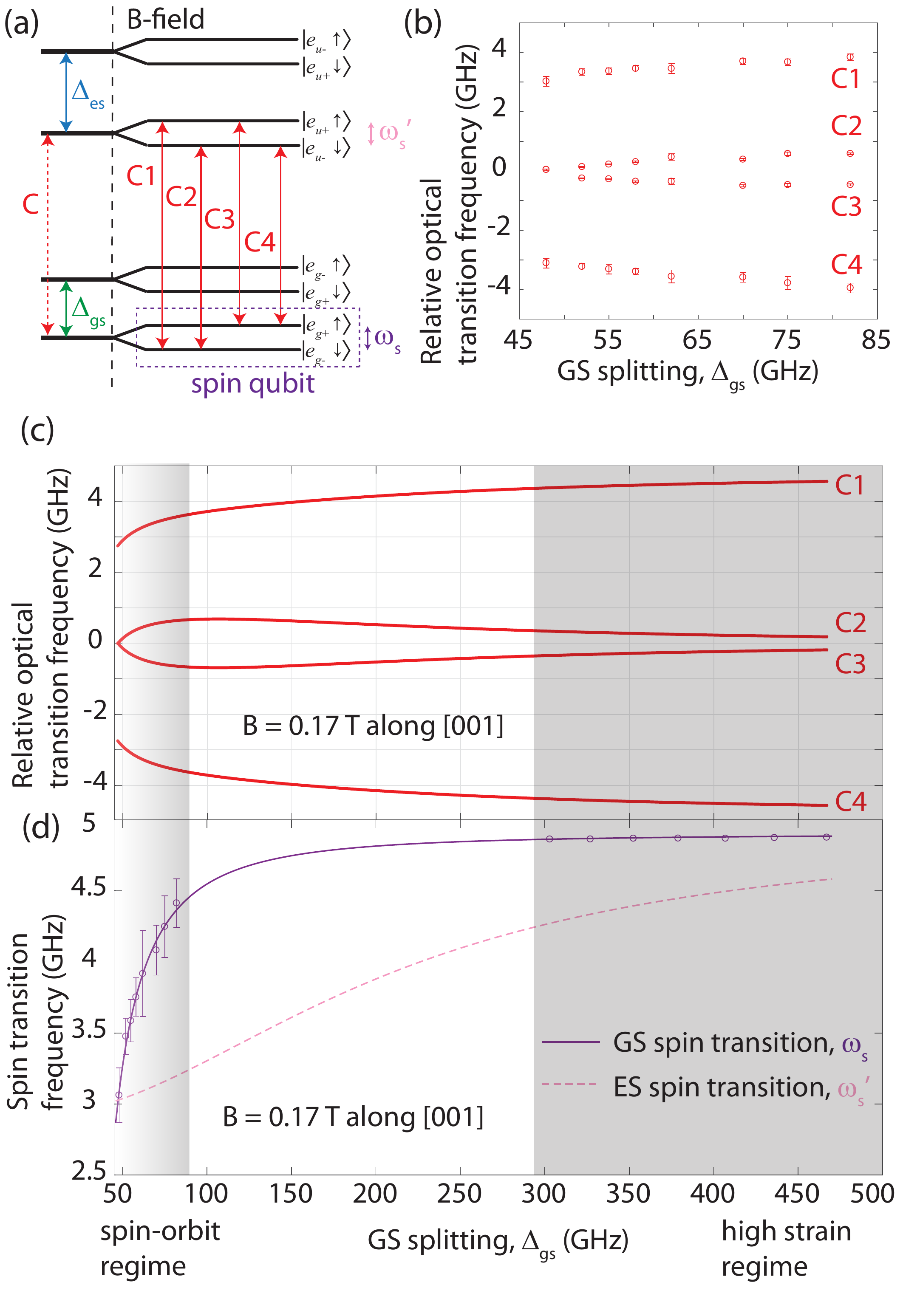}
\caption{(a) Splitting of the C transition into the four transitions C1, C2, C3, and C4 in the presence of a magnetic field. Spin transition frequencies on the lower orbital branches of the GS and ES are $\omega_s$, $\omega_s^\prime$ respectively. (b) Response of transitions C1, C2, C3, and C4 upon tuning GS splitting $\Delta_{\mathrm{gs}}$ with $E_g$-strain. (c) Calculated response of optical transitions C1, C2, C3, and C4 to $E_g$-strain in presence of 0.17 T B-field aligned along the [001] direction. Shaded regions on the left and right ends indicate the regimes in which the GS orbitals are determined by SO coupling and strain respectively. (d) Strain response of spin transition frequencies upon tuning of ground state orbital splitting $\Delta_{\mathrm{gs}}$ with $E_g$-strain. SO regime data points are extracted from the optical spectra in Fig. 5(b). High strain regime data points are obtained from CPT measurements on the SiV studied in Fig. 4. Solid (dashed) line is calculated spin transition frequency on the lower orbital branch of GS (ES) from Fig. 5(c).}
\label{fig6}
\end{figure}

So far, we have seen that static $E_g$-strain in the SiV environment can significantly impact spin coherence and relaxation rates by modifying the orbital splitting in the GS. In this section, we discuss additional effects of this type of strain on the SiV spin qubit that arise from SO coupling. Particularly, we can tune the spin transition frequency, $\omega_s$ by a large amount (a few GHz) at a fixed external magnetic field by simply controlling local strain. At the same time, we discuss how the magnitude of local strain strongly determines the ability to couple or control the SiV spin qubit with external fields such as resonant strain or microwaves at frequency $\omega_s$, and resonant laser-fields in a $\Lambda$-scheme.

The strain-response of the spin transition is measured by monitoring the four Zeeman-split optical lines arising from the C transition as shown schematically in Fig. \ref{fig6}(a). In Fig. \ref{fig6}(b), we apply a fixed magnetic field  $B=$0.17 T aligned along the vertical [001] axis with a permanent magnet placed underneath the sample, and gradually increase the GS splitting of a transverse-orientation SiV by applying strain. With increasing strain, each of the four Zeeman-split optical transitions moves outwards from the position of the unsplit C transition at zero magnetic field. In particular, the spin-conserving inner transitions C2 and C3 overlap at zero strain, but become more resolvable with increasing strain. Thus, all-optical control of the spin \cite{becker2017all} relying on simultaneous excitation of a pair of transitions C1 and C3 (or C2 and C4) forming a $\Lambda$-scheme requires the presence of some local strain. The strain-tuning behavior of Zeeman split optical transitions can be theoretically calculated by diagonalizing the GS and ES Hamiltonians in the presence of a magnetic field. Upon adding Zeeman terms to the Hamiltonian in equation \ref{eq:H.SO.strain}, and switching to the basis of SO eigenstates $\{e_{g-}\downarrow, e_{g+}\uparrow, e_{g+}\downarrow, e_{g-}\uparrow \}$, we obtain

\begin{widetext}
\begin{equation}
\mathbb{H}^{\text{total}} = \begin{bmatrix} 
-\lambda_{SO}/2-\gamma_LB_z-\gamma_sB_z & 0 & \epsilon_{E_{gx}} & \gamma_sB_x \\
0 & -\lambda_{SO}/2+\gamma_LB_z+\gamma_sB_z & \gamma_sB_x & \epsilon_{E_{gx}} \\
\epsilon_{E_{gx}} & \gamma_sB_x & \lambda_{SO}/2+\gamma_LB_z-\gamma_sB_z & 0\\
\gamma_sB_x & \epsilon_{E_{gx}} & 0 & \lambda_{SO}/2-\gamma_LB_z+\gamma_sB_z
\end{bmatrix}
\label{eq:Htot.lowstr}
\end{equation}
\end{widetext}

Here we have discarded the $A_{1g}$ and $E_{gy}$ strain terms, since the transverse-orientation SiVs in our experiments experience predominantly $E_{gx}$ strain. We have also assumed that the transverse magnetic field is entirely along the $x$-axis of the SiV. The gyromagnetic ratios are $\gamma_s$ = 14 GHz/T, $\gamma_L$ = 0.1(14) GHz/T, where the pre-factor of 0.1 is a quenching factor for the orbital angular momentum.\cite{HeppThesis} The result of our calculation is shown in Fig. \ref{fig6}(c). In the low strain regime indicated by the region with the shaded gradient, we reproduce the experimental behavior in Fig. \ref{fig6}(b), and obtain good quantitative agreement with the variation in the spin transition frequency $\omega_s$ in Fig. \ref{fig6}(d). 

Physically, this behavior of the spin transitions arises as strain and SO coupling compete to determine the orbital wavefunctions. From the Hamiltonian in equation \ref{eq:Htot.lowstr}, we can see that the orbitals begin as SO eigenstates $\{e_{g-}\downarrow, e_{g+}\uparrow, e_{g+}\downarrow, e_{g-}\uparrow \}$ at zero strain, and end up as the pure states $\{e_{gx}\downarrow, e_{gx}\uparrow, e_{gy}\downarrow, e_{gy}\uparrow \}$ at high strain ($\epsilon_{E_{gx}} \gg \lambda_{SO}/2$). At zero strain, the effective magnetic field from SO coupling quantizes the electron spin along the $z-$axis. In this condition, the off-axis B-field does not affect the spin transition frequency $\omega_s$ to first order, so $\omega_s \sim 2(\gamma_s+\gamma_L)B_z = 3.1$ GHz. As the strain $\epsilon_{E_{gx}}$ is increased far above the SO coupling $\lambda_{SO}$ and the orbitals are purified, the spin quantization axis approaches the direction of the external magnetic field, and $\omega_s$ approaches ~$2\gamma_sB = 4.8$ GHz. Since SO coupling in the ES is stronger, this limit is attained at higher values of strain than in the GS as shown by the dashed line in Fig. \ref{fig6}(d). Once the orbitals in both the GS and ES are predominantly dictated by local strain and SO coupling is merely perturbative, the difference in GS and ES spin transition frequencies becomes vanishingly small, eventually leading to converging C2 and C3 optical transitions as depicted on the right hand side of Fig. \ref{fig6}(c). In the limit of very high strain, the transitions C2 and C3 also become strictly spin-conserving, and optical polarization \cite{rogers_electronic_2014, pingault_coherent_2017} and readout of the spin qubit will be forbidden.

The rapid variation of the spin transition frequency $\omega_s$ in the low-strain regime of Fig. \ref{fig6}(d) provides the first hint that the SiV spin-qubit can be very sensitive to oscillating strain generated by coherent phonons. The interaction terms due to strain and the off-axis magnetic field predicted by the Hamiltonian in equation \ref{eq:Htot.lowstr} are depicted visually in Fig. \ref{fig7}(a). In particular, at zero strain, the presence of the off-axis magnetic field perturbs the eigenstates of the spin qubit to first order as
\begin{eqnarray}
\ket{e_{g-}\downarrow}^\prime & \approx & \ket{e_{g-}\downarrow} + \frac{\gamma_sB_x}{\lambda_{SO}}\ket{e_{g-}\uparrow} \label{eq:spinstates.perturbed1}\\
\ket{e_{g+}\uparrow}^\prime & \approx & \ket{e_{g+}\uparrow} + \frac{\gamma_sB_x}{\lambda_{SO}}\ket{e_{g+}\downarrow} \label{eq:spinstates.perturbed2}
\end{eqnarray}

This perturbative mixing with opposite spin-character can now allow resonant AC strain at frequency $\omega_s$ to drive the spin qubit. For a small amplitude of such AC strain $\epsilon_{E_{gx}}^{AC}$, we can calculate the strain susceptibility of the spin transition $d_{\mathrm{spin}}$ in terms of the GS orbital strain susceptibility $d_g$ in Eq. \ref{eq:str.susc.vals}.

\begin{equation}
d_{\mathrm{spin}} = \frac{\bra{e_{g-}\downarrow^\prime}\mathbb{H}^{\text{strain}}\ket{e_{g+}\uparrow^\prime}}{\epsilon_{E_{gx}}^{AC}}d_g = \frac{2\gamma_sB_x}{\lambda_{SO}}d_g
\label{eq:dspin.pert} 
\end{equation}

Since $d_g$ is very large ($\sim$1 PHz/strain), even with the presence of the pre-factor ${\gamma_sB_x}/{\lambda_{SO}}$, the spin qubit can have a relatively large strain-response. For the present case of $B$=0.17 T along the [001] axis, we get $d_{\mathrm{spin}}/d_\perp = 0.085$ yielding $d_{\mathrm{spin}} \sim 100$ THz/strain. An exact calculation of $d_{\mathrm{spin}}$ for arbitrary local static strain using the Hamiltonian in equation \ref{eq:Htot.lowstr} is shown in Fig. \ref{fig7}(b). As static strain in the SiV environment is increased far above the SO coupling, the AC strain susceptibility approaches zero. Thus we can conclude that coupling the SiV spin qubit to resonant AC strain requires (i) low static strain $\epsilon_{E_{g}} \ll \lambda_{SO}/2$ and (ii) a non-zero off-axis magnetic field $B_x$. The spin qubit can also parametrically couple to off-resonant AC strain with a different susceptibility $t_{\mathrm{spin}}$, and this is discussed in Appendix \ref{t.spin}. A similar analysis predicts the response of the spin qubit to resonant microwave magnetic fields in Appendix \ref{microwaves}.

\section{Prospects for a coherent spin-phonon interface}
\begin{figure}[h!]
\includegraphics[width=\columnwidth]{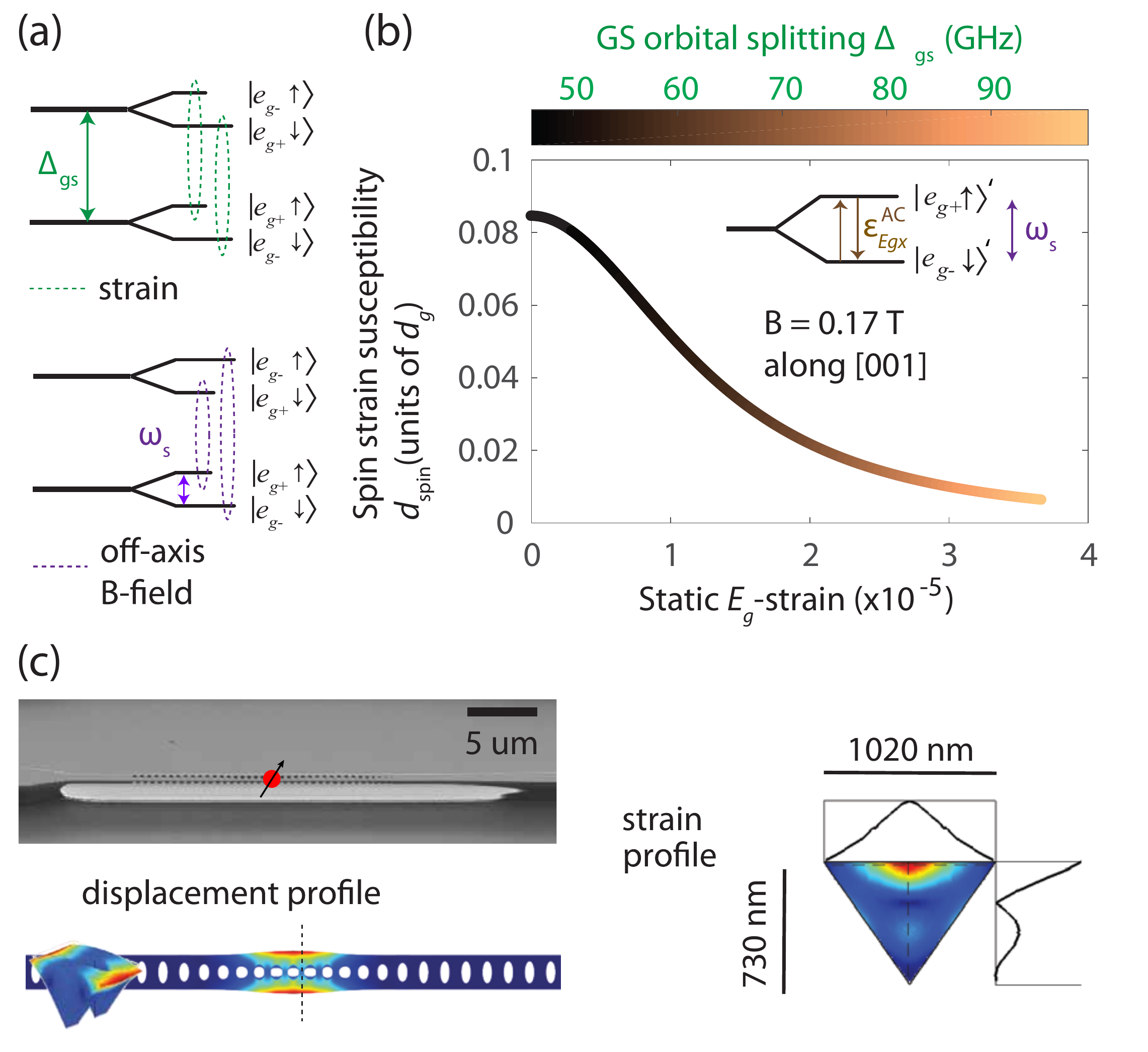}
\caption{(a) Illustration of mixing terms introduced by $E_g$-strain and an off-axis magnetic field in the GS manifold. (b) Calculated susceptibility of the spin-qubit for interaction with AC $E_g$-strain resonant with the transition frequency $\omega_s$ (interaction shown in inset). This AC strain susceptibility is maximum at zero strain for the pure SO eigenstates. At high strain, it falls off as $1/\Delta_{\mathrm{gs}}$. Color variation along the curve shows the GS splitting $\Delta_{\mathrm{gs}}$ corresponding to the value of static $E_g$-strain at the SiV. Both the static and AC strain are assumed to be entirely in the $\beta$ component. (c) SEM image of an OMC nanobeam cavity \cite{burek_diamond_2016} along with an FEM simulation of its 5 GHz flapping resonance. Displacement profile and a cross-sectional strain profile of the mode are shown with arbitrary normalization.}
\label{fig7}
\end{figure}

Our results on the strain response of the electronic and spin levels of the SiV indicate the potential of this color center as a spin-phonon interface. The diamond NV center spin, the most investigated candidate in this direction has an intrinsically weak strain susceptibility ($\sim 10$ GHz/strain) since the qubit levels are defined within the same orbital in the GS configuration of the defect\cite{barson2017nanomechanical}. While using distinct orbitals in the ES can provide much larger strain susceptibility ($\sim 1$ PHz/strain)\cite{davies_optical_1976, lee_strain_2016}, such schemes will be limited by fast dephasing due to spontaneous emission and spectral diffusion. In comparison, the SiV center provides distinct orbital branches within the GS itself. Further, the presence of SO coupling dictates that the spin qubit levels $\ket{e_{g-}\downarrow}, \ket{e_{g+}\uparrow}$ correspond to different orbitals. As a result, one achieves the ideal combination of high strain susceptibility and low qubit dephasing rate.

The effects of various modes of strain and the rich electronic structure of the SiV allow a variety of spin-phonon coupling schemes. In this letter, we focus on direct coupling of the spin transition to a mechanical resonator at frequency $\omega_s$ enabled by $E_g$-strain response of the spin discussed in the previous section. An alternative approach utilizing propagating phonons of frequency $\sim \lambda_\mathrm{SO}$ coupled to the GS orbital transition is discussed elsewhere\cite{lemonde_2017}. Our scheme would require diamond mechanical resonators of frequency $\omega_s \sim$ few GHz, which have already been realized in both optomechanical\cite{burek_diamond_2016,lake_2017} and electromechanical platforms\cite{macquarrie_coherent_2015,macquarrie_continuous_2015,golter_optomechanical_2016,golter_saw_2016}. Fig. \ref{fig7}(c) shows the strain profile resulting from GHz frequency mechanical modes in an optomechanical crystal cavity. Since this structure achieves three-dimensional confinement of phonons on the scale of the acoustic wavelength, it provides large per-phonon strain. For an SiV located $\sim$20 nm below the top surface, when a magnetic field $B =$ 0.3 T is applied along the [001] direction, the spin qubit is resonant with the 5 GHz flapping mode, and has a single-phonon coupling rate $g \sim$0.8 MHz. At mK temperatures, given the low SiV spin dephasing rate $\gamma_s \sim$ 100 Hz\cite{sukachev2017silicon}, even modest mechanical quality-factors $Q_m\sim 10^3$ measured previously\cite{burek_diamond_2016} are sufficient to achieve strong spin-phonon coupling. At 4 K, despite the higher spin dephasing rate $\gamma_s \sim$ 4 MHz\cite{pingault_all-optical_2014,rogers_all-optical_2014} and thermal occupation of mechanical modes $n_{th} \sim 20$, high spin-phonon co-operativity can be achieved if previously observed 4 K quality factors for silicon OMCs\cite{chan_2012}, $Q_m \sim 10^5$ can be replicated in diamond. This form of spin-phonon coupling can also be implemented in other resonator designs such as surface acoustic wave cavities\cite{golter_optomechanical_2016,golter_saw_2016,lee_topical_2017}, wherein piezoelectric materials are used to transduce the mechanical motion with microwave electrical signals instead of optical fields.

\section{Conclusion}
In conclusion, we characterize the strain response of the SiV center in diamond with a NEMS device. The implications of our results are two-fold. First, the large tuning range of optical transitions we have demonstrated establishes strain control as a technique to achieve spectrally identical emitters in a quantum network. Strain tuning is particularly relevant here since inversion-symmetric centers with superior optical properties do not have a first order electric field response, thereby negating the feasibility of direct electrical tuning. Second, the intrinsic sensitivity of the SiV spin qubit to strain makes it a promising candidate for coherent spin-phonon coupling. This can enable phonon-mediated quantum information processing with spins\cite{wallquist_hybrid_2009,rabl_quantum_2010}. The development of such a cavity QED platform with a phononic two-level system\cite{ruskov2013chip,aref_2015} will also allow deterministic quantum nonlinearities for phonons\cite{oconnell_quantum_2010}, thereby overcoming inefficiencies in probabilistic schemes used to generate single phonon states in cavity optomechanics\cite{riedinger2016,hong2017}. Further, the use of optomechanical and electromechanical resonators towards this goal suggests the possibility of coherently interfacing diamond spin qubits with telecom and microwave photons respectively.

\appendix
\section{Group theoretical description of strain response} \label{group.theory}
The response of the electronic levels of trigonal point-defects in cubic crystals to lattice deformations was treated theoretically by Hughs and Runciman\cite{hughes_uniaxial_1967}. A solution of this problem for the specific case of the SiV has been previously carried out using group theory \cite{HeppThesis} with some errors. Here, we reconcile these two treatments, and present a model for the response of the SiV electronic levels to strain (and stress). In what follows, we use $x, y, z$ to refer to the internal basis of the SiV (see inset of Fig. \ref{fig1}(b). eg. for a $[111]$ oriented SiV, we have $x:[\bar{1}\bar{1}2], y:[\bar{1}10], z:[111]$), and $X,Y,Z$ to refer to the axes of the diamond crystal, i.e. $X:[100], Y:[010], Z:[001]$. We use $\sigma$ and $\epsilon$ for the stress and strain tensors in the SiV basis, and $\bar{\sigma}$ and $\bar{\epsilon}$ to refer to them in the crystal basis. We also neglect the spin character of the states involved, since we are only concerned with changes to the Coulomb energy of the orbitals.

When the applied stress is small, in the Born-Oppenheimer approximation, the effect of lattice deformation is linear in the strain components and is captured by a Hamiltonian of the form\cite{hughes_uniaxial_1967} -

\begin{equation}
\mathbb{H}^{\text{strain}} = \sum_{ij}{A_{ij} {\epsilon}_{ij}}
\label{eq:HstrGeneral1}
\end{equation}

Here $i, j$ are indices for the co-ordinate axes. $V_{ij}$ are operators corresponding to particular stress components, and act on the SiV electronic levels. Group theory can be used to rewrite this Hamiltonian in terms of basis-independent linear combinations of strain components adapted to the symmetries of the SiV center. Each of these combinations can be viewed as a particular `mode' of deformation, and the effect of each mode on the orbital wavefunctions, each with its own symmetries can be deduced using group theory. More technically, such deformation modes are obtained by projecting the strain tensor onto the irreducible representations of $D_{3d}$, the point group of the SiV center.\cite{hughes_uniaxial_1967} This transformation gives 

\begin{equation}
\mathbb{H}^{\text{strain}} = \sum_{r} V_{r}\epsilon_{r} 
\label{eq:HstrGeneral2}
\end{equation}

where $r$ runs over the irreducible representations. Deducing the operators $V_{r}$ simply requires computing the direct products of irreducible representations.\cite{HeppThesis} It can be shown that strain and stress tensors transform as the irreducible representation,$A_{1g} + E_g$\cite{HeppThesis} which has even parity about the inversion center of the SiV. Since the ground states of the SiV transform as $E_g$ (even), and the excited states transform as $E_u$ (odd), lattice deformations do not couple the ground and excited states with each other to first order. As a result, we can describe the response of the ground and excited state manifolds independently. In particular, $\mathbb{H}^{\text{strain}}$ is identical in form for both manifolds, but will involve different numerical values of strain-response coefficients. Therefore, we drop the subscripts $g$ and $u$ used to refer to the ground and excited states, and simply work in the doubly-degenerate basis $\{\ket{e_x},\ket{e_y}\}$. The interaction Hamiltonian can be shown to comprise three deformation modes -

\begin{equation}
\mathbb{H}^{\text{strain}} = 
\alpha
\begin{bmatrix}
1 & 0  \\
0 & 1  \\
\end{bmatrix} + \beta
\begin{bmatrix}
-1 & 0  \\
0 & 1  \\
\end{bmatrix} + \gamma
\begin{bmatrix}
0 & 1  \\
1 & 0  \\
\end{bmatrix}
\label{eq:Hstr}
\end{equation}
\\
The components $\alpha, \beta, \gamma$ corresponding to $\epsilon_r$ in Eq.\ref{eq:HstrGeneral2} are given by the following linear combinations\cite{hughes_uniaxial_1967}

\begin{eqnarray}
\alpha &=& \mathscr{A}_1(\bar{\epsilon}_{XX}+\bar{\epsilon}_{YY}+\bar{\epsilon}_{ZZ}) + 2\mathscr{A}_2(\bar{\epsilon}_{YZ}+\bar{\epsilon}_{ZX}+\bar{\epsilon}_{XY}) \nonumber\\
\beta &=& \mathscr{B}(2\bar{\epsilon}_{ZZ}-\bar{\epsilon}_{XX}-\bar{\epsilon}_{YY}) + \mathscr{C}(2\bar{\epsilon}_{XY}-\bar{\epsilon}_{YZ}-\bar{\epsilon}_{ZX}) \nonumber\\
\gamma &=& \sqrt{3}\mathscr{B}(\bar{\epsilon}_{XX}-\bar{\epsilon}_{YY}) + \sqrt{3}\mathscr{C}(\bar{\epsilon}_{YZ}-\bar{\epsilon}_{ZX}) \nonumber
\end{eqnarray}

The coefficients $\mathscr{A}_1, \mathscr{A}_2, \mathscr{B}, \mathscr{C}$ completely determine the strain-response of the $\{\ket{e_x},\ket{e_y}\}$ manifold. It can be shown that $\alpha$ transforms as $A_{1g}$, and $\{\beta,\gamma\}$ transform as $\{E_{gx},E_{gy}\}$.

To gain more physical intuition for these three deformation modes, we can write $\alpha,\beta,\gamma$ in the SiV basis using the unitary transformation $R = R_z(45^\circ)R_y(54.7^\circ)$, where $R_z(\theta)$, and $R_x(\phi)$ correspond to rotations by $\theta$ and $\phi$ about the $z$- and $x$-axes respectively. Upon transformation, we get

\begin{eqnarray}
\alpha & = & t_\perp(\epsilon_{xx}+\epsilon_{yy}) + t_\parallel\epsilon_{zz} \equiv \epsilon_{A_{1g}} \nonumber\\
\beta & = & d(\epsilon_{xx}-\epsilon_{yy}) + f\epsilon_{zx} \equiv \epsilon_{E_{gx}} \label{eq:strain.siv.basis}\\
\gamma & = & -2d\epsilon_{xy} + f\epsilon_{yz} \equiv \epsilon_{E_{gy}} \nonumber
\end{eqnarray}

Here $t_\perp, t_\parallel, d, f$ are the four strain-susceptibility parameters. They are related to the original stress-response coefficients of Hughs and Runciman\cite{hughes_uniaxial_1967} according to the expressions in Table \ref{tab:strain.modes}. Further, to explicitly indicate the symmetries of these deformation modes, we hereafter switch to the notation $\epsilon_{A_{1g}}$ for $\alpha$, $\epsilon_{E_{gx}}$ for $\beta$, and $\epsilon_{E_{gy}}$ for $\gamma$ in line with the description in Eq. (\ref{eq:HstrGeneral2}).

At this juncture, we contrast Eqs. (\ref{eq:strain.siv.basis}) with the recent results in Ref. \cite{HeppThesis} (Eqs. 2.80-2.82). Our analysis predicts a non-zero response to uniaxial strain along the high symmetry axis $\epsilon_{zz}$ in $A_{1g}$ deformation, and to the shear strains $\epsilon_{zx}$ and $\epsilon_{yz}$ in $E_g$ deformations.


\begin{widetext}
\begin{table*}
\centering
\begin{tabular}{|c|c|c|}
\hline
Strain term & Susceptibility & Relation to Hughes-Runciman coefficients \\
\hline
${\epsilon}_{xx}+{\epsilon}_{yy}$ & $t_{\perp}$ & $(c_{11}+2c_{12})A_1 - c_{44}A_2$ \\
\hline
${\epsilon}_{zz}$ & $t_{\parallel}$ & $(c_{11}+2c_{12})A_1 + 2c_{44}A_2$ \\
\hline
${\epsilon}_{xx}-{\epsilon}_{yy}$ & $d$ & $(c_{11}-c_{12})B + c_{44}C$ \\
\hline
${\epsilon}_{xy}$ & $-2d$ & \\
\hline
${\epsilon}_{zx}$ & $f$ & $\sqrt{2}\left(c_{44}C - 2(c_{11}-c_{12})B\right)$ \\
\hline
${\epsilon}_{yz}$ & $f$ & \\
\hline
\end{tabular}
\caption{Various strain-modes, and their susceptibilities in terms of the Hughs-Runciman stress-response coefficients\cite{hughes_uniaxial_1967}. The constants $c_{ij}$ are the elastic modulus components of diamond.}
\label{tab:strain.modes}
\end{table*}
\end{widetext}

\section{Extraction of strain susceptibilities}\label{strain.susc.fit}
To extract all the values $\{t_\perp, t_\parallel, d, f\}$ for both ground and excited state manifolds, in principle, strain needs to be applied at least in three different directions for a given SiV. This procedure gives a set of overdetermined equations in these parameters.\cite{hughes_uniaxial_1967} However, the devices in this study can only induce two types of strain profiles as shown in Fig. \ref{fig3}(c) and (d). In particular, for a given SiV in either the `axial' or the `transverse' class, the relative ratio between strain-tensor components remains constant, when voltage applied to the cantilever is swept. This condition makes it difficult to estimate the relative contributions of $t_\parallel$ and $t_\perp$ to $\epsilon_{A_{1g}}$, and of $d$ and $f$ to $\epsilon_{E_{g}}$.\\

To get around this issue, we follow an approximate approach. From Fig. \ref{fig3}(d), we observe that in the case of an axial SiV, $\epsilon_{zz} \gg \left(\epsilon_{xx} + \epsilon_{yy}\right)$ is always true. Therefore, we can use the response of the axial SiV in Fig. \ref{fig2}(b) to approximately estimate $\left(t_{\parallel,u}-t_{\parallel,g}\right)$ by neglecting $\left(\epsilon_{xx} + \epsilon_{yy}\right)$ in Eq. (\ref{eq:fitmodel1}). Fig. \ref{fig3}(f) plots the mean ZPL frequency of the axial SiV in Fig. \ref{fig2}(b) vs. $\epsilon_{zz}$ estimated from FEM simulation. The slope of the linear fit yields $\left(t_{\parallel,u}-t_{\parallel,g}\right)$.

\begin{equation}
\left(t_{\parallel,u}-t_{\parallel,g}\right) = -1.7\,\text{PHz/strain}
\end{equation}

Likewise, in the case of the transverse SiV in Fig. \ref{fig3}(c), we can conclude that $\left(\epsilon_{xx}-\epsilon_{yy}\right) \gg \text{max}\{\epsilon_{zx}, \epsilon_{yz}\}$. With this class of SiVs, we can approximately estimate $\{d_g, d_u\}$ by neglecting $\{\epsilon_{zx}, \epsilon_{yz}\}$ in Eqs. (\ref{eq:fitmodel2},\ref{eq:fitmodel3}). Fig. \ref{fig3}(e) plots the GS and ES splittings of the transverse SiV in Fig. \ref{fig2}(a) vs. $\epsilon_\perp = \sqrt{\left(\epsilon_{xx}-\epsilon_{yy}\right)^2 + 4\epsilon_{xy}^2}$ estimated from FEM simulation. Fitting yields

\begin{equation}
\qquad d_g = 1.3,
\qquad d_u = 1.8
\qquad \text{PHz/strain}
\label{eq:estimation1}
\end{equation}

Once we extract $\left(t_{\parallel,u}-t_{\parallel,g}\right)$ from an axial SiV, we can use this value to further extract $\left(t_{\perp,u}-t_{\perp,g}\right)$ by fitting Eq. (\ref{eq:fitmodel1}) to the tuning behavior of the mean ZPL frequency of the transverse SiV. This procedure yields

\begin{equation}
\left(t_{\perp,u}-t_{\perp,g}\right)=78\,\text{THz/strain}
\label{eq:estimation2}
\end{equation}

We immediately note that $\left(t_{\parallel,u}-t_{\parallel,g}\right)$ is more than an order of magnitude larger than $\left(t_{\perp,u}-t_{\perp,g}\right)$. This implies that $\epsilon_{zz}$ tunes the mean ZPL frequency much more effectively than $(\epsilon_{xx}+\epsilon_{yy})$. This can be intuitively explained by examining the spatial profile of the GS and ES orbitals (Table 2.7 of Ref. \cite{HeppThesis}). Since the GS and ES correspond to even ($g$) and odd ($u$) eigenstates of SiV's $D_{3d}$ point symmetry group respectively, the charge density distributions of the orbitals $e_{gx},e_{ux}$ (and $e_{gy},e_{uy}$) are similar in any transverse plane normal to the $z$-axis. As a result, we would expect that the common mode energy shift resulting from the strain-mode $\epsilon_{xx}+\epsilon_{yy}$ is very similar for the GS and ES manifolds, i.e. $t_{\perp,u}\approx t_{\perp,g}$. On the other hand, the energy shift from $\epsilon_{zz}$ is expected to have opposite signs for the GS and ES manifolds due to the change in wavefunction parity along the $z$-axis. \\

As the last step, we extract the values $f_g, f_u$ in Eqs. (\ref{eq:fitmodel2},\ref{eq:fitmodel3}). We observe from table \ref{tab:strain.modes} that knowledge of $d$ and $B$ can allow us to determine $f$. The Hughs-Runciman coefficients $B_g$=484\,GHz/GPa and $B_u$=630\,GHz/GPa can be extracted based on uniaxial stress measurements carried out in Ref. \cite{sternschulte_1.681-ev_1994,HeppThesis}. Combining our estimates of $d_g$ and $d_u$ with this information, we predict

\begin{equation}
f_g=-250,
\qquad f_u=-720
\qquad \text{THz/strain}
\label{eq:estimation3}
\end{equation}

\section{Spin relaxation ($T_1$) model}\label{spin.T1}
\begin{figure}[!ht]
\centering
\includegraphics[width=\columnwidth]{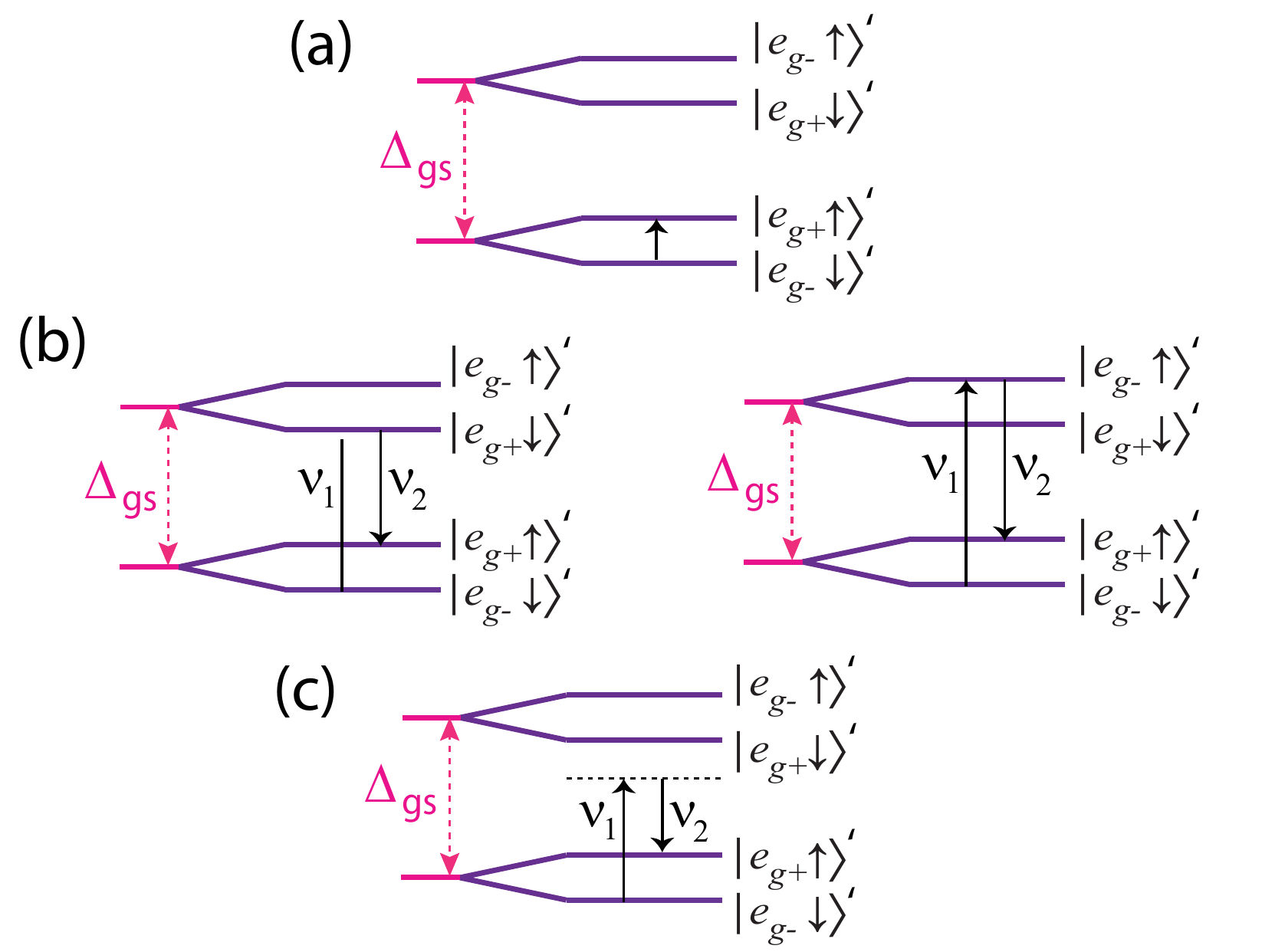}
\caption{Various pathways for a phonon-mediated spin-flip (a) Direct relaxation via a single phonon resonant with the $\ket{e_{g-}\downarrow}^\prime \rightarrow \ket{e_{g+}^\uparrow}\prime$ spin-transition. (b) Two possible channels for a resonant two-phonon process involving the upper orbital branch. (c) Off-resonant two-phonon processes.}%
\label{SpinT1Processes}%
\end{figure}
$E_g$-phonons predominantly drive spin-conserving transitions between the GS orbitals of the SiV i.e. between $\{\ket{e_{g-}\downarrow}^\prime, \ket{e_{g+}\downarrow}^\prime\}$, and $\{\ket{e_{g+}\uparrow}^\prime, \ket{e_{g-}\uparrow}^\prime\}$ respectively. However, in the presence of an off-axis magnetic-field, and non-zero static strain, the eigenstates of the GS manifold are no longer pure SO or strain eigenstates, and all transitions between the four states within the GS manifold become allowed for $E_g$-phonons. In this scenario, the various channels for spin-relaxation from $\ket{e_{g-}\downarrow}^\prime$ to $\ket{e_{g+}\uparrow}^\prime$ are:

\begin{itemize}
	\item Direct single-phonon relaxation: Via a single phonon of frequency $\omega_{\mathrm{s}}$ resonant with the spin-transition as shown in Fig. \ref{SpinT1Processes}(a)
	\item Resonant two-phonon relaxation: Via two phonons resonant with a level in the upper orbital branch as an intermediate state as shown in Fig. \ref{SpinT1Processes}(b). The spin-flip can be caused by either the emitted phonon (left) or the absorbed phonon (right).
	\item Off-resonant two-phonon relaxation: Via two phonons with a virtual level as an intermediate state as shown in Fig. \ref{SpinT1Processes}(c). The effective driving strength will be reduced from its value in the resonant process by an amount corresponding to the detuning from the upper orbital branch.
\end{itemize}

\begin{figure}%
\centering
\includegraphics[width=\columnwidth]{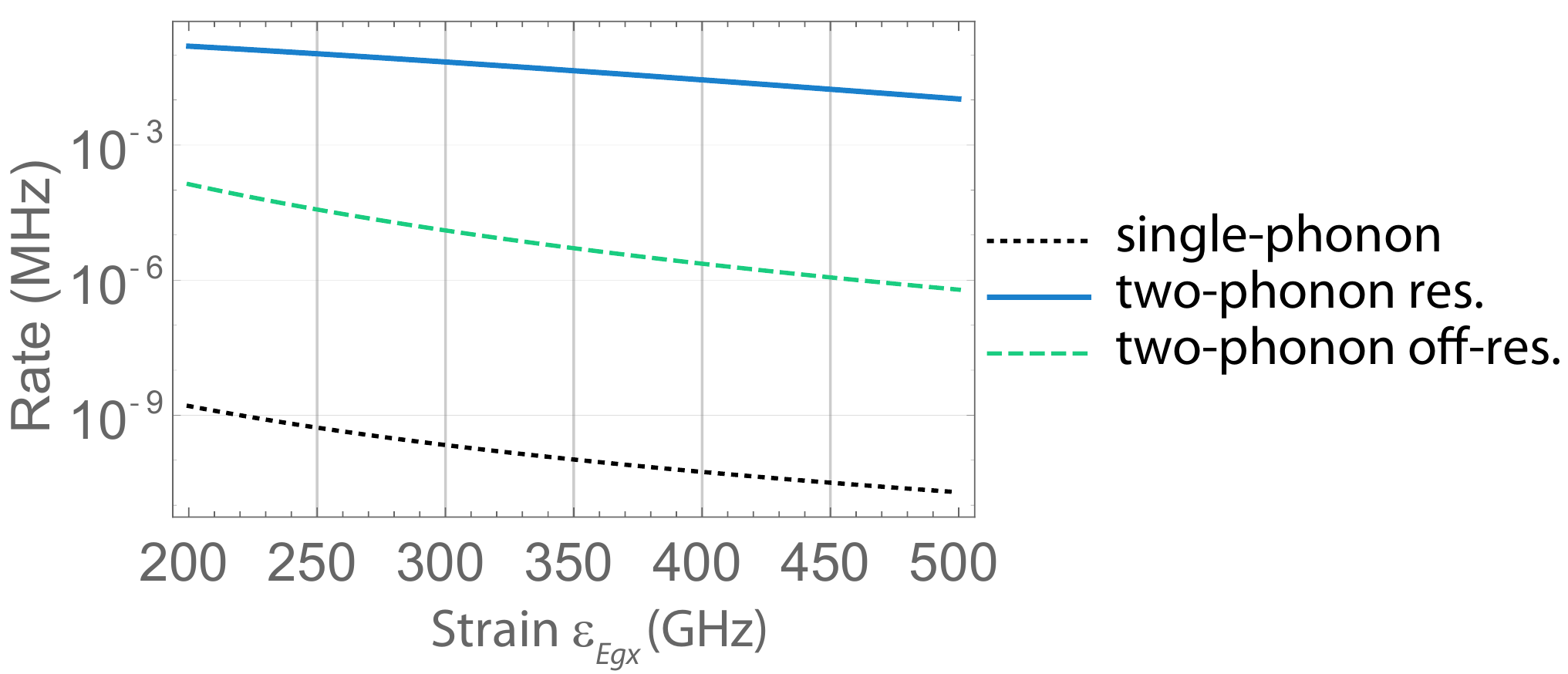}%
\caption{Rates of all three spin-relaxation mechanisms indicating their magnitudes and scaling with strain $\beta$.}%
\label{spin.relax.summary}%
\end{figure}

Using Fermi's golden rule, the transition rates for these relaxation channels can be calculated. The results are summarized in Table \ref{tab:spin.relaxation}, and are plotted versus GS splitting $\Delta_\mathrm{gs}$ in Fig. \ref{spin.relax.summary}. 

\begin{table*}[t!]
\centering
\begin{tabular}{|c|c|c|c|}
\hline
Mechanism & Rate & Relevant regime & Expected scaling of rate \\
\hline
Single-phonon & $2\pi \left(\frac{d_\mathrm{spin}}{d_g}\right)^2\chi\rho\omega_s^3 n_{th}(\omega_s)$ & $k_BT/h \ll \omega_s$ & $B_\perp^2\Delta_{\mathrm{gs}}^{-2}\omega_s^3 \mathrm{exp}(-h\omega_s/k_BT)$ \\
\hline
Resonant two-phonon & $4 \left(\frac{d_{g, \mathrm{flip}}}{d_g}\right)^2 \gamma_{\mathrm{up}} $ & $k_BT/h \sim \Delta_{\mathrm{gs}}$ & $B_\perp^2\Delta_{\mathrm{gs}}[\mathrm{exp}(h\Delta_{\mathrm{gs}}/k_BT)-1]^{-1}$ \\
\hline
Off-resonant two-phonon & $8\pi^3 \left(\frac{d_{g, \mathrm{flip}}}{d_g}\right)^2 \chi^2\rho^2 {\omega_s^2} \left(\frac{k_BT}{h}\right)^3$ & $k_BT/h \gg \Delta_{\mathrm{gs}}$ & $B_\perp^2\Delta_{\mathrm{gs}}^{-2}\omega_s^2T^3$ \\
\hline
\end{tabular}
\caption{Summary of spin-relaxation mechanisms}
\label{tab:spin.relaxation}
\end{table*}

We see that spin relaxation at 4 K is dominated by a two-phonon process involving the upper ground state orbital branches as intermediate states. In literature, this is frequently referred to as an Orbach process.\cite{orbach_spin-lattice_1961} The experimentally observed behavior of spin $T_1$ in Fig. \ref{fig5}(d) of the main text is well-explained by the scaling of such a process with the GS splitting $\Delta_{\mathrm{gs}}$ shown in Table \ref{tab:spin.relaxation}. Intuitively, we may understand the dominance of the Orbach process in terms of the phonon DOS $\propto \Delta^n \mathrm{exp}\left(-h\Delta/k_BT\right)$ being maximized around the frequency $\Delta\sim k_BT/h$. We can similarly argue that the single and off-resonant two-phonon channels become relevant in other temperature regimes indicated in Table \ref{tab:spin.relaxation}, where the phonon DOS is maximized in a frequency range relevant for those processes. 

\section{Dispersive strain-coupling to spin qubit} \label{t.spin}
\begin{figure}%
\centering
\includegraphics[width=\columnwidth]{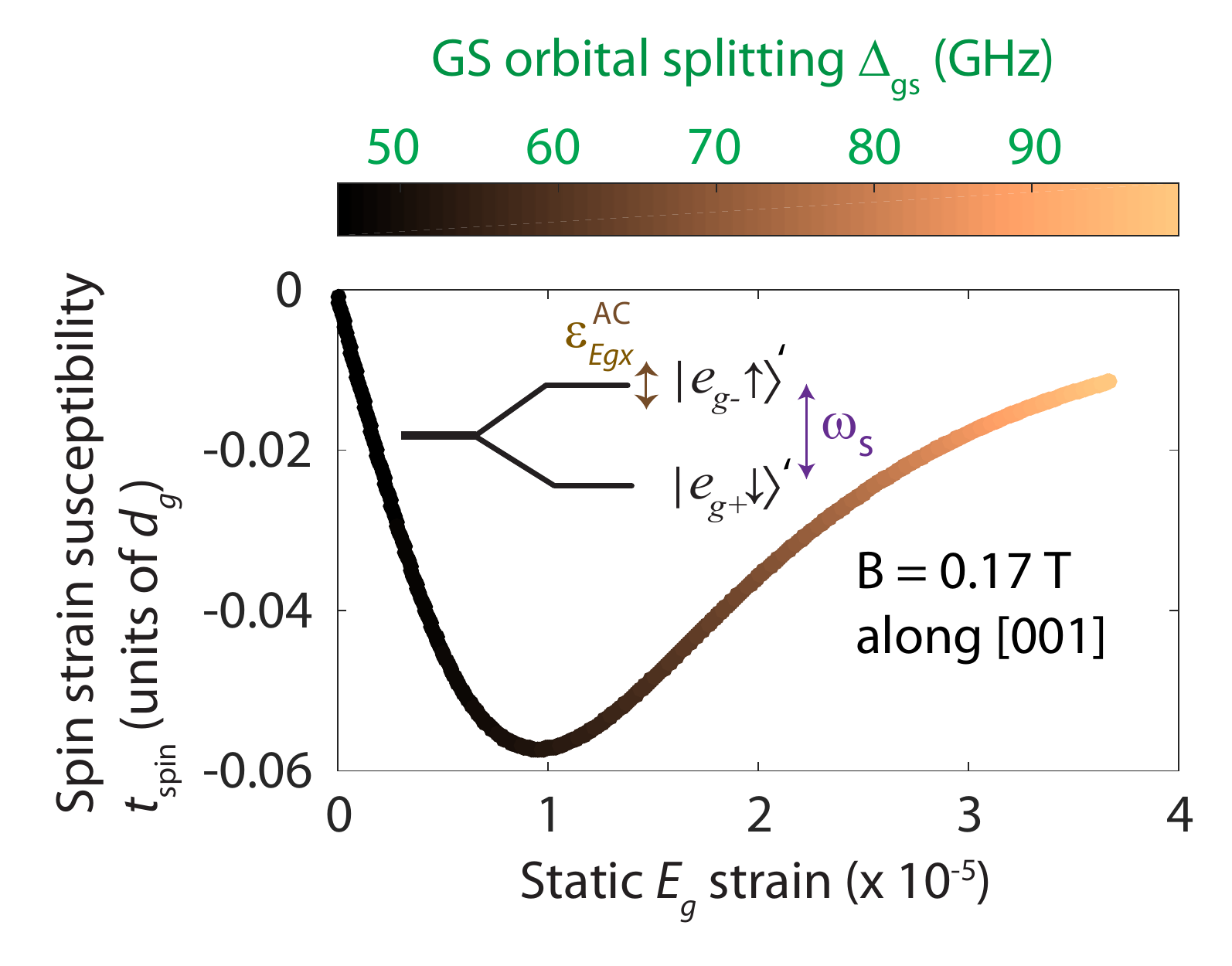}
\caption{Calculated susceptibility of the spin-qubit for interaction with off-resonant AC $E_g$-strain that modulates the transition frequency $\omega_s$ (interaction shown in inset). Color variation along the curve shows the GS splitting corresponding to the value of static $E_g$-strain at the SiV. Both the DC and AC strain are assumed to be entirely in the $E_{gx}$-component.}%
\label{tspin.vs.strain}%
\end{figure}
From Eqs. (\ref{eq:spinstates.perturbed1}, \ref{eq:spinstates.perturbed2}), we concluded that in the the low strain limit, the eigenstates of the SiV spin qubit $\ket{e_{g-}\downarrow}^\prime$, $\ket{e_{g+}\uparrow}^\prime$ are linearly mixed by $E_g$-strain, and hence suitable for resonant driving by AC strain at frequency $\omega_s$. This type of mixing also indicates that static $E_g$-strain would cause a quadratic shift in the spin-transition frequency $\omega_s$. Such a quadratic response to an external field can always generate a linear AC response in the presence of a `bias' field. Thus in the presence of non-zero static $E_g$-strain, $\omega_s$ must also experience a linear modulation with off-resonant AC strain. This is particularly useful for parametric coupling of the spin qubit to off-resonant mechanical resonators as demonstrated previously with NV centers\cite{ovartchaiyapong_dynamic_2014, teissier_2014, barfuss_strong_2015, meesala_enhanced_2016}. A calculation of the magnitude of modulation in the spin transition frequency for a given AC strain $\beta_{AC}$ yields the susceptibility $t_{spin}$ for dispersive spin-phonon coupling, which can be of the same order of magnitude as $d_{spin}$.

\begin{equation}
t_{spin} = \frac{\bra{e_{g+}\uparrow^\prime}\mathbb{H}_{str}^{AC}\ket{e_{g+}\uparrow^\prime} - \bra{e_{g-}\downarrow^\prime}\mathbb{H}_{str}^{AC}\ket{e_{g-}\downarrow^\prime}}{\beta_{AC}}d_g
\label{eq:tspin}
\end{equation}

$t_{spin}$ is calculated as a function of pre-existing static $E_g$-strain, and plotted in Fig. \ref{tspin.vs.strain}. Its magnitude is maximized at a moderately strained GS splitting of 50 GHz, and falls off as static strain is further increased. This non-monotonic behavior arises from the fact that $t_{spin}$ is a result of linearizing the quadratic response due to $d_{spin}$, and therefore scales as the product of $d_{spin}$ and static strain in the environment. Thus there is an optimal static strain condition to maximize $t_{spin}$.

\section{Microwave magnetic response of spin qubit} \label{microwaves}

\begin{figure}[t]
\centering
\includegraphics[width=\columnwidth]{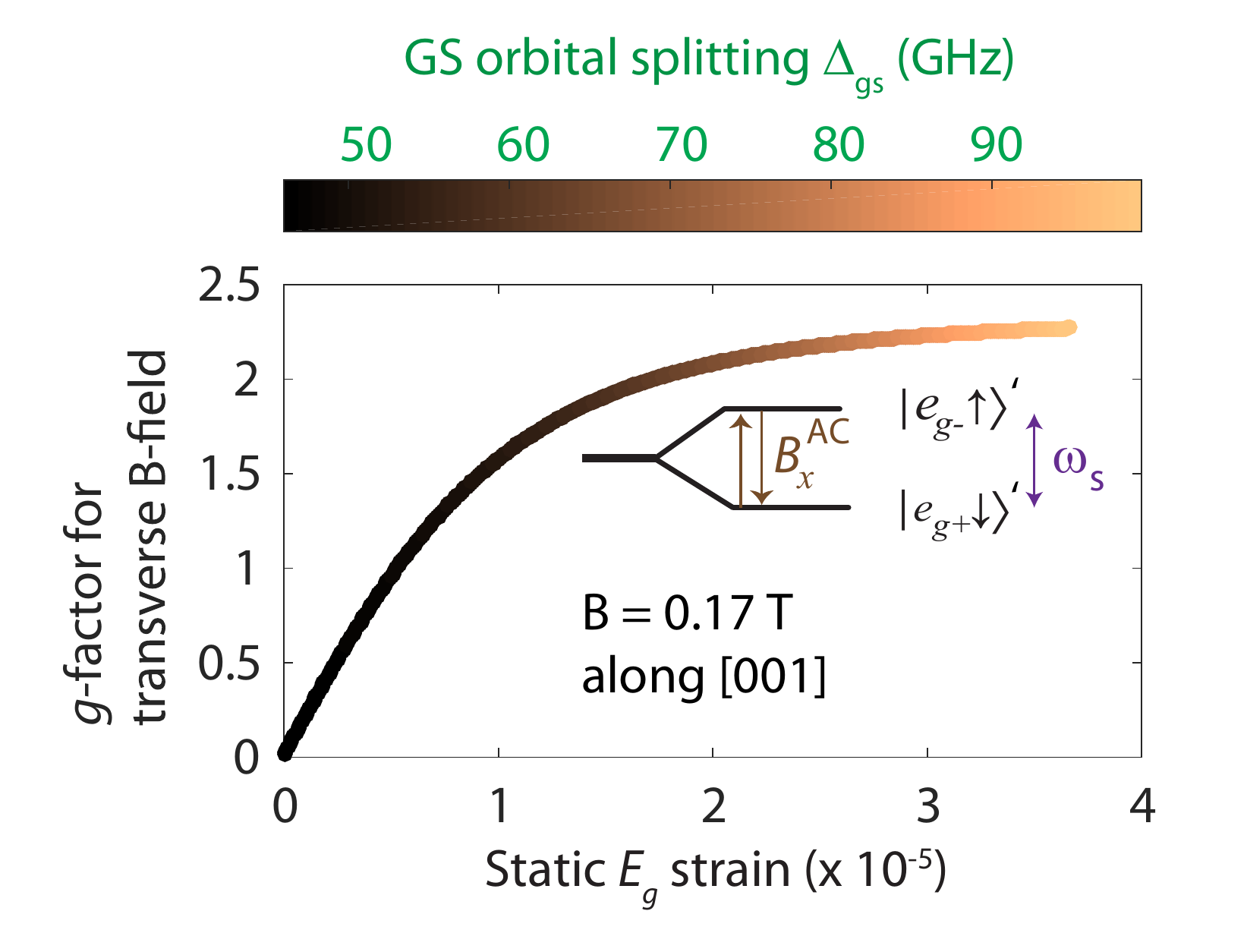}
\caption{$g$-factor for a transverse microwave magnetic field resonantly driving the spin qubit (interaction shown in inset) calculated as a function of pre-existing static $E_g$-strain plotted along $x-$ axis. Color variation along the curve shows the GS splitting corresponding to the value of static $E_g$-strain at the SiV. Both the DC and AC strain are assumed to be entirely in the $E_{gx}$-component.}%
\label{gFactor.vs.strain}%
\end{figure}

At zero strain, the spin qubit cannot be driven by microwave magnetic fields at frequency $\omega_s$. This is because a magnetic field cannot flip the orbital character of the pure SO eigenstates $\ket{e_{g-}\downarrow}$, $\ket{e_{g+}\uparrow}$ that comprise the spin qubit as evinced by the Hamiltonian \ref{eq:Htot.lowstr}. However, just as a transverse B-field allows a strain susceptibility for the spin qubit as shown by Eqs. (\ref{eq:spinstates.perturbed1}-\ref{eq:dspin.pert}), we can argue that the presence of strain induces a non-zero response to transverse B-fields. Fig. \ref{gFactor.vs.strain} shows a calculation of the effective $g$-factor for transverse B-field $B_{x}^{AC}$, which determines the Rabi frequency $\Omega_{MW} = g_x\mu_B B_{x}^{AC}$ for microwave control\cite{pingault_coherent_2017} of the SiV spin. As expected, the $g$-factor approaches close to that of a free electron at high strain, when the SO coupling can be neglected.

\begin{acknowledgments}
This work was supported by STC Center for Integrated Quantum Materials (NSF Grant No. DMR-1231319), ONR MURI on Quantum Optomechanics (Award No. N00014-15-1-2761), NSF EFRI ACQUIRE (Award No. 5710004174), the University of Cambridge, the ERC Consolidator Grant PHOENICS, the EPSRC Quantum Technology Hub NQIT (EP/M013243/1), and the MIT-Harvard CUA. B.P. thanks Wolfson College (University of Cambridge) for support through a research fellowship. Device fabrication was performed in part at the Center for Nanoscale Systems (CNS), a member of the National Nanotechnology Infrastructure Network (NNIN), which is supported by the National Science Foundation under NSF award no. ECS-0335765. CNS is part of Harvard University. Focused ion beam implantation was performed under the Laboratory Directed Research and Development Program at the Center for Integrated Nanotechnologies, an Office of Science User Facility operated for the U.S. Department of Energy (DOE) Office of Science. Sandia National Laboratories is a multi-mission laboratory managed and operated by National Technology and Engineering Solutions of Sandia, LLC., a wholly owned subsidiary of Honeywell International, Inc., for the U.S. Department of Energy's National Nuclear Security Administration under contract DE-NA-0003525. We thank D. Perry for performing the focused ion beam implantation, and M. W. Doherty for helpful discussions.
\end{acknowledgments}

\end{document}